\magnification 1200
\baselineskip 15 pt

\def \m {\mu}
\def \s {\sigma}

\def \an {A^{(n)}}

\def \l {\lambda}
\def \la {\lambda}
\def \a {\alpha}
\def \ul {\underline \lambda}

\def \P {\Phi}
\def \Pn {\Phi^{(n)}}

\def \bn {B^{(n)}}

\def \sect#1{\bigskip  \noindent{\bf #1} \medskip }
\def \subsect#1{\bigskip \noindent{\it #1} \medskip}
\def \subsubsect #1{\bigskip \noindent{ #1} \medskip}
\def \th#1#2{\medskip \noindent {\bf Theorem #1.}   \it #2 \rm \medskip}

\def \cor#1#2{\medskip \noindent {\bf Corollary #1.}   \it #2 \rm \medskip}
\def \pf {\noindent  {\it Proof}.\quad }
\def \lem#1#2{\medskip \noindent {\bf Lemma #1.}   \it #2 \rm \medskip}

\def\sqr#1#2{{\vcenter{\vbox{\hrule height.#2pt\hbox{\vrule width.#2pt height#1pt \kern#1pt\vrule width.#2pt}\hrule height.#2pt}}}}

\def \square{\hfill\mathchoice\sqr56\sqr56\sqr{4.1}5\sqr{3.5}5}

\centerline{\bf Pricing Life Insurance under Stochastic Mortality} \medskip
\centerline{\bf via the Instantaneous Sharpe Ratio: Theorems and Proofs} \bigskip

\centerline {Version: 10 May 2007}
\bigskip

\medskip

\noindent Virginia R. Young \hfill \break
\indent Department of Mathematics \hfill \break
\indent University of Michigan \hfill \break
\indent Ann Arbor, Michigan, 48109 \hfill \break
\indent vryoung@umich.edu

\bigskip \bigskip

\noindent{\bf Abstract:}  We develop a pricing rule for life insurance under stochastic mortality in an incomplete market by assuming that the insurance company requires compensation for its risk in the form of a pre-specified instantaneous Sharpe ratio.  Our valuation formula satisfies a number of desirable properties, many of which it shares with the standard deviation premium principle.  The major result of the paper is that the price per contract solves a {\it linear} partial differential equation as the number of contracts approaches infinity.  One can interpret the limiting price as an expectation with respect to an equivalent martingale measure.   Another important result is that if the hazard rate is stochastic, then the risk-adjusted premium is greater than the net premium, even as the number of contracts approaches infinity.  We present a numerical example to illustrate our results, along with the corresponding algorithms.

\bigskip


\bigskip


\sect{1. Introduction}

We propose a pricing rule for life insurance when interest rates and mortality rates are stochastic by applying the method developed and expounded upon by Milevsky, Promislow, and Young (2005, 2007).  In the case addressed in this paper, their method amounts to targeting a pre-specified Sharpe ratio for a portfolio of bonds that optimally hedges the life insurance, albeit only partially.

Actuaries often assume that one can eliminate the uncertainty associated with mortality by selling a large number of insurance contracts.  This assumption is valid if the force of mortality is deterministic.  Indeed, if the insurer sells enough contracts, then the deviation of actual results from what is expected goes to zero, so the risk is diversifiable.  However, because the insurer can only sell a finite number of insurance policies, it is impossible to eliminate the risk that experience will differ from what is expected.  The risk associated with selling a finite number of insurance contracts is what we call the {\it finite portfolio risk}.

On the other hand, if the force of mortality for a population is stochastic, then there is a systematic (that is, common) risk that cannot be eliminated by selling more policies.  We call this risk the {\it stochastic mortality risk}, a special case of stochastic parameter risk.  Even as the insurer sells an arbitrarily large number of contracts, the systematic stochastic mortality risk remains.

We argue that mortality is uncertain and that this uncertainty is correlated across individuals in a population -- mostly due to medical breakthroughs or environmental factors that affect the entire population.  For example, if there is a positive probability that medical science will find a cure for cancer during the next thirty years, this will influence aggregate mortality patterns.  Biffis (2005), Schrager (2006), Dahl (2004), as well as Milevsky and Promislow (2001), use diffusion processes to model the force of mortality, as we do in this paper.  One could model catastrophic events that affect mortality widely, such as epidemics, by allowing for random jumps in the force of mortality.  Also see the related work of Cox and Lin (2004) and Cairns, Blake, and Dowd (2004). 

In related work, Blanchett-Scalliet, El Karoui, and Martellini (2005) value assets that mature at a random time by using the principle of no arbitrage by focusing on equivalent martingale measures; the resulting pricing rules are, therefore, linear.  Dahl and M\o ller (2006) take a similar approach in their work.  However, for insurance markets, one cannot assert that no arbitrage holds, so we use a different method to value life insurance contracts and our resulting pricing rule is non-linear, except in the limit. 

We value life insurance by assuming that the insurance company is compensated for its risk via the so-called {\it instantaneous Sharpe ratio} of a suitably-defined portfolio.  Specifically, we assume that the insurance company picks a target ratio of expected excess return to standard deviation, denoted by $\a$, and then determines a price for a life insurance contract that yields the given $\a$ for the corresponding portfolio.  One might call $\a$ the market price of mortality risk, but it appears in the pricing equation in a non-linear manner.  However, as the number of life insurance policies increases to infinity, then this $\a$ is the market price of risk in the ``traditional'' sense of pricing in financial markets in that it acts to modify the drift of the hazard rate process.

In this paper, we assume that the life insurance company does not sell annuities to hedge the stochastic mortality risk.  In related work, Bayraktar and Young (2007a) allow the insurer to hedge its risk partially by selling pure endowments to individuals whose stochastic mortality is correlated with that of the buyers of life insurance. 

We obtain a number of results from our methodology that one expects within the context of insurance.  For example, we prove that if the hazard rate is deterministic, then as the number of contracts approaches infinity, the price of life insurance converges to the net premium under the physical probability; see Corollary 4.4.  In other words, if the stochastic mortality risk is not present, then the price for a large number of life insurance policies reflects this and reduces to the usual expected value pricing rule in the limit.

An important theorem of this paper is that as the number of contracts approaches infinity, the limiting price per risk solves a {\it linear} partial differential equation and can be represented as an expectation with respect to an equivalent martingale measure as in Blanchett-Scalliet, El Karoui, and Martellini (2005); see Corollary 4.6.  Therefore, we obtain their results as a limiting case of ours.  Moreover, if the hazard rate is stochastic, then the value of the life insurance contract is greater than the net premium, even as the number of contracts approaches infinity; see Corollary 4.5.

The remainder of this paper is organized as follows. In Section 2, we present our financial market, describe how to use the instantaneous Sharpe ratio to price life insurance payable at the moment of death, and derive the resulting partial differential equation (pde) that the price $A = A^{(1)}$ solves.  We also present the pde for the price $\an$ of $n$ conditionally independent and identically distributed life insurance risks.   In Section 3, we study properties of $\an$; our valuation operator is subadditive and satisfies a number of other appealing properties.  In Section 4, we find the limiting value of ${1 \over n} \an$ and show that it solves a linear pde.  We also decompose the risk charge for a portfolio of life insurance policies into a systematic component (due to uncertain aggregate mortality) and a non-systematic component (due to insuring a finite number of policies); see equation (4.18).  In Section 5, we present a numerical example that illustrates our results, along with the corresponding algorithms that we use in the computation.  Section 6 concludes the paper.

\sect{2. Instantaneous Sharpe Ratio}

In this section, we describe the term life insurance policy and present the financial market in which the issuer of this contract invests.  We obtain the hedging strategy for the issuer of the life insurance.  We describe how to use the instantaneous Sharpe ratio to price the insurance policy, and we derive the resulting partial differential equation (pde) that the price solves; see equation (2.16).  We also present the pde for the price $\an$ of $n$ conditionally independent and identically distributed life insurance risks; see equation (2.18).

\subsect{2.1  Mortality Model and Financial Market}

We use the stochastic model of mortality of Milevsky,  Promislow, and Young (2005).  We model the hazard rate for an individual or set of individuals of a given age.  If we were to model a population's hazard rate, then we would take into account age and time as in Lee and Carter (1992) and, more recently, Ballotta and Haberman (2006).  However, because we consider a single age, we simply model the hazard rate as a stochastic process over time.

We assume that the hazard rate $\la_t$ (or force of mortality) of an individual at time $t$ follows a diffusion process such that if the process begins at $\la_0 > \ul$ for some positive constant $\ul$, then $\la_t > \ul$ for all $t \ge 0$.  From a modeling standpoint, $\ul$ could represent the lowest attainable hazard rate remaining after all causes of death such as accidents and homicide have been eliminated; see, for example, Gavrilov and Gavrilova (1991) and Olshansky, Carnes, and Cassel (1990).

Specifically, we assume that

$$d\la_t = \mu(\la_t, t) (\la_t - \ul) dt + \sigma(t) (\la_t - \ul) dW^\la_t, \eqno(2.1)$$

\noindent in which $\{W^\la_t \}$ is a standard Brownian motion on a probability space $(\Omega, {\cal F}, {\bf P})$.  The volatility $\s$ is either identically zero, or it is a continuous function of time $t$ bounded below by a positive constant $\kappa$ on $[0, T]$.  The drift $\m$ is a H\"older continuous function of $\la$ and $t$ for which there exists $\epsilon > 0$ such that if $0 < \la - \ul < \epsilon$, then $\mu(\la, t) > 0$ for all $t \in [0, T]$.  After Lemma 3.2 below, we add more requirements for $\mu$.  Note that if $\s \equiv 0$, then $\la_t$ is deterministic, and in this case, we write $\l(t)$ to denote the deterministic hazard rate at time $t$.

Suppose an insurer issues a term life insurance policy to an individual that pays \$1 at the moment of death if the individual dies before some time $T$.  In Section 2.2, to determine the value of the insurance policy, we create a portfolio composed of the obligation to underwrite the life insurance, of default-free, zero-coupon bonds that pay \$1 at time $T$, regardless of the state of the individual, and money invested in a money market account earning at the short rate.  Therefore, we require a model for bond prices, and we use a model based on the short rate and the bond market's price of risk.

The dynamics of the short rate $\{r_t \}$, which is the rate at which the money market increases, are given by

$$dr_t = b(r_t, t) dt + d(r_t, t) dW_t, \eqno(2.2)$$

\noindent in which $b$ and $d \ge 0$ are deterministic functions of the short rate and time, and $\{W_t \}$ is a standard Brownian motion with respect to the probability space $(\Omega, {\cal F}, {\bf P})$, independent of $\{W^\la_t \}$.  We assume that $b$ and $d \ge 0$ are such that $r_t \ge 0$ almost surely for all $t \ge 0$.  Assuming that the dynamics of the short rate and the hazard rate are independent is reasonable because one does not necessarily expect changes in interest rates to affect changes in mortality rates and vice versa.  However, if one believes that it is more accurate to allow the two processes to be correlated, then the work in this paper extends appropriately, albeit with added complexity because of the correlaton.

From the principle of no-arbitrage in the bond market, there is a market price of risk process $\{ q_t \}$ for the bond that is adapted to the filtration generated by $\{W_t \}$; see, for example, Lamberton and Lapeyre (1996) or Bj\"ork (2004).  Moreover, the bond market's price of risk at time $t$ is a deterministic function of the short rate and of time, that is, $q_t = q(r_t, t)$.  Thus, the time-$t$ price $F$ of a default-free, zero-coupon bond that pays \$1 at time $T$ is given by

$$F(r, t; T) = {\bf E^Q} \left[e^{-\int_t^T r_s ds} ds \Bigg | r_t = r \right], \eqno(2.3)$$

\noindent in which $\bf Q$ is the probability measure with Radon-Nikodym derivative with respect to $\bf P$ given by ${d{\bf Q} \over d{\bf P}} \big|_{{\cal F}_t} = \exp \left({-\int_0^t q(r_s, s) dW_s - {1 \over 2} \int_0^t q^2(r_s, s) ds} \right)$.  It follows that $\{W^Q_t \}$, with $W^Q_t = W_t + \int_0^t q(r_s, s) ds$, is a standard Brownian motion with respect to $\bf Q$.

From Bj\"ork (2004), we know that the $T$-bond price $F$ solves the following pde:

$$\left\{ \eqalign{& F_t + b^Q(r, t) F_r + {1 \over 2} d^2(r, t) F_{rr} - r F = 0, \cr
&F(r, T; T) = 1,} \right. \eqno(2.4)$$

\noindent in which  $b^Q = b - q d$.  Henceforth, we drop the notational dependence of $F$ on $T$ because $T$ is fixed (and understood) throughout the paper.

We can use the pde (2.4) to obtain the dynamics of the $T$-bond price $F(r_s, s)$, in which we think of $r_t = r$ as given and $s \in [t, T]$.  Indeed,

$$\left\{ \eqalign{dF(r_s, s) &= (r_s \, F(r_s, s) + q(r_s, s) \, d(r_s, s) \, F_r(r_s, s)) ds + d(r_s, s) \, F_r(r_s, s) \, dW_s, \cr
F(r_t, t) &= F(r, t).} \right. \eqno(2.5)$$

\noindent We use (2.5) in the next section when we develop the dynamics of a portfolio containing the obligation to pay the life insurance and a certain number of $T$-bonds.

\subsect{2.2  Pricing via the Instantaneous Sharpe Ratio}

In this section, we first describe our method for pricing in this (incomplete) life insurance market.  Then, we fully develop the pricing rule for a single life insurance policy; that is, $n = 1$.  Next, we consider some special cases when $n = 1$.  Finally, we present the pricing rule for the case $n \ge 1$.

\subsubsect{2.2.1 Recipe for valuation}

The market for insurance is incomplete; therefore, there is no unique pricing mechanism.  To value contracts in this market, one must assume something about how risk is ``priced.''  For example, one could use the principle of equivalent utility (see Zariphopoulou (2001) for a review) or consider the set of equivalent martingale measures (Dahl and M\o ller, 2006) to price the risk.  We use the instantaneous Sharpe ratio because of its analogy with the bond market's price of risk and because of the desirable properties of the resulting price; see Section 3.  Also, in the limit as the number of contracts approaches infinity, the resulting price can be represented as an expectation with respect to an equivalent martingale measure; see Corollary 4.6 below.  Because of these properties, we anticipate that our pricing methodology will prove useful in pricing risks in other incomplete markets.

Our method for pricing in an incomplete market is as follows:

\item{A.}  First, define a portfolio composed of two parts:  (1) the obligation to underwrite the contingent claim (life insurance, in this case), and (2) a self-financing sub-portfolio of $T$-bonds and money market funds to (partially) hedge the contingent claim.

\item{B.}  Find the investment strategy in the $T$-bonds so that the local variance of the total portfolio is a minimum.  If the market were complete, then one could find an investment strategy so that the local variance is zero.  However, in an incomplete market, there will be residual risk as measured by the local variance.  This control of the local variance is related to local risk-minimization as described in Schweizer (2001).

\item{C.}  Determine the price of the contingent claim so that the instantaneous Sharpe ratio of the total portfolio equals a pre-specified value.  This amounts to setting the drift of the portfolio equal to the short rate times the portfolio value {\it plus} the Sharpe ratio time the local standard deviation of the portfolio.  Bj\"ork and Slinko (2006) apply the idea of limiting the instantaneous Sharpe ratio to restrict the possible prices of claims in an incomplete market.

\subsubsect{2.2.2 Pricing a single life insurance policy}

Denote the time-$t$ value (price) of a term life insurance policy that pays \$1 at the time of death of the individual between times $t$ and $T$ by $A = A(r, \la, t)$, in which we explicitly recognize that the price of the insurance depends on the short rate $r$ and the hazard rate $\la$ at time $t$.  (As an aside, by writing $A$ to represent the value of the life insurance policy, we mean that the buyer of the policy is still alive.  If the individual dies before time $T$, then the value of the policy jumps to \$1.)

Suppose the insurer creates a portfolio with value $\Pi_t$ at time $t$.  The portfolio contains (1) the obligation to underwrite the life insurance of \$1, with value $-A$, and (2) a self-financing sub-portfolio of $T$-bonds and money market funds with value $V_t$ at time $t$ to (partially) hedge the risk of the insurance policy.  Thus, $\Pi_t = -A(r_t, \la_t, t) + V_t$.  Let $\pi_t$ denote the number of $T$-bonds in the self-financing sub-portfolio at time $t$, with the remainder, namely $V_t - \pi_t F(r_t, t)$, invested in a money market account earning the short rate $r_t$.

The insurance risk cannot be fully hedged because of the randomness inherent in the individual living or dying.  If the individual dies at time $t < T$, then the value of the insurance policy jumps to \$1 because the insurer is obligated to pay \$1 immediately; therefore, the value of the portfolio $\Pi_t$ jumps from $-A(r, \la, t)  + V_t$ to $-1 + V_t$.  In other words, the value of the portfolio changes instantly changes by $A(r, \la, t) - 1$, or equivalently, the portfolio value decreases by $1 - A(r, \la, t)$, the so-called {\it net amount at risk} (Bowers et al., 1997).

We next describe the dynamics of the total portfolio $\Pi_t$ by specifying the dynamics of its two pieces $A(r_t, \la_t, t)$ and $V_t$.  By It\^o's Lemma (see, for example, Protter (1995), the dynamics of the value of the life insurance policy are given as follows:

$$\eqalign{dA(r_t, \la_t, t) &= A_t \, dt + A_r \, dr_t + {1 \over 2} A_{rr} \, d[r, r]_t + A_\la \, d\la_t + {1 \over 2} A_{\la \la} \, d[\la, \la]_t + (A - 1) \, dN_t, \cr
&= A_t \, dt + A_r (b \, dt + d \, dW_t) + {1 \over 2} A_{rr} \, d^2 \, dt + A_\la (\la_t - \ul) (\mu \, dt + \sigma \, dW^\la_t) \cr
& \quad + {1 \over 2} A_{\la \la} \, \sigma^2 (\la_t - \ul)^2 dt + (A - 1) \, dN_t,} \eqno(2.6)$$

\noindent in which we suppress the dependence of the functions $a$, $b$, etc. on the variable $r$, $\la$, and $t$.  Also, $[r, r]$, for example, denotes the quadratic variation of $r$, and $\{N_t\}$ denotes a counting process with stochastic parameter $\la_t$ at time $t$ that indicates when the individual dies.  Because $V_t = \pi_t F(r_t, t) + (V_t - \pi_t F(r_t, t))$ is self-financing, its dynamics are given by

$$\eqalign{dV_t &= \pi_t \, dF(r_t, t) + r_t (V_t - \pi_t F(r_t, t)) dt = (\pi_t \, q \, d \, F_r + r_t \, V_t) \, dt + \pi_t \, d \, F_r \, dW_t,} \eqno(2.7)$$

\noindent in which the second equality follows from (2.5) and we again suppress the dependence of the functions on the underlying variables.

It follows from equations (2.6) and (2.7) that the value of the portfolio at time $t + h$ with $h > 0$, namely $\Pi_{t + h}$, equals

$$\eqalign{\Pi_{t+h} &= \Pi_t - \int_t^{t+h} dA(r_s, \la_s, s) + \int_t^{t+h} dV_s \cr
&= \Pi_t - \int_t^{t+h} {\cal D}^b A(r_s, \la_s, s) ds + \int_t^{t+h} d(r_s, s) (\pi_s F_r(r_s, s) - A_r(r_s, \la_s, s)) \, dW_s \cr
& - \int_t^{t+h} \s(s) (\la_s - \ul) A_\la(r_s, \la_s, s) \, dW^\la_s + \int_t^{t+h} (A(r_s, \la_s, s) - 1) (dN_s - \la_s \, ds)  \cr
& + \int_t^{t+h}  \pi_s \, q(r_s, s) \, d(r_s, s) \, F_r(r_s, s) \, ds + \int_t^{t+h} r_s \, \Pi_s \, ds,} \eqno(2.8)$$

\noindent in which ${\cal D}^b$ is an operator defined on the set of appropriately differentiable functions on ${\bf R}^+ \times (\ul, \infty) \times [0, T]$ by

$${\cal D}^b v = v_t + b \, v_r + {1 \over 2} d^2 \, v_{rr} + \m (\la - \ul) v_\la + {1 \over 2} \s^2 (\la - \ul)^2 \, v_{\la \la} - \la (v - 1) - rv.  \eqno(2.9)$$

\noindent  Note that we adjust $dN_t$ in (2.8) so that $N_t - \int_0^t \la_s \, ds$ is a martingale.

In this single-life case, the process $\{ \Pi_t \}$ is ``killed'' when the individual dies.  If we were to consider the price $\an$ of $n$ conditionally independent and identically distributed lives (conditionally independent given the hazard rate), then $\{ N_t \}$ would be a counting process with stochastic parameter $n \la_t$ at time $t$ such that $\Pi_t$ decreases by $1 - \left(\an  - A^{(n - 1)} \right)$ when one of the $n$ individuals dies.  We consider $\an$ later in this section and continue with the single-life case now.

We choose $\pi_t$ to minimize the local variance of this portfolio.  To this end, we calculate the expectation and variance of $\Pi_{t+h}$ conditional on the information available at time $t$, namely ${\cal F}_t$.  First, given $\Pi_t = \Pi$, define a stochastic process $Y_h$ for $h \ge 0$ by

$$Y_h = \Pi - \int_t^{t+h} {\cal D}^b A(r_s, \la_s, s) ds + \int_t^{t+h} \pi_s \, q(r_s, s) \, d(r_s, s) \, F_r(r_s, s) \, ds + \int_t^{t+h} r_s \, \Pi_s \, ds.  \eqno(2.10)$$

\noindent Thus, ${\bf E}(\Pi_{t+h} | {\cal F}_t) = {\bf E}^{r, \la, t} (Y_h)$, in which ${\bf E}^{r, \la, t}$ denotes the conditional expectation given $r_t = r$ and $\la_t = \la$.  Also, from (2.8) and (2.10), we have

$$\eqalign{&\Pi_{t + h} = Y_h + \int_t^{t+h} d(r_s, s) (\pi_s F_r(r_s, s) - A_r(r_s, \la_s, s)) \, dW_s \cr
& \quad - \int_t^{t+h} \s(s) (\la_s - \ul) A_\la(r_s, \la_s, s) \, dW^\la_s + \int_t^{t+h} (A(r_s, \la_s, s) - 1) (dN_s - \la_s \, ds).} \eqno(2.11)$$

\noindent It follows that

$$\eqalign{& {\bf Var}(\Pi_{t+h} | {\cal F}_t) = {\bf E}((\Pi_{t+h} - {\bf E}Y_h)^2 | {\cal F}_t) \cr
& = {\bf E}^{r, \la, t} (Y_h - {\bf E}Y_h)^2 + {\bf E}^{r, \la, t} \int_t^{t+h} d^2(r_s, s) (\pi_s F_r(r_s, s) - A_r(r_s, \la_s, s))^2 ds \cr
&+ {\bf E}^{r, \la, t}  \int_t^{t+h} \s^2(s) (\la_s - \ul)^2 A_\la^2(r_s, \la_s, s) ds + {\bf E}^{r, \la, t}  \int_t^{t+h} \la_s (A(r_s, \la_s, s) - 1)^2 ds.} \eqno(2.12)$$

Recall that we wish to choose $\pi_t$ to minimize the local variance $\lim_{h \rightarrow 0} {1 \over h} {\bf Var}(\Pi_{t+h} | {\cal F}_t)$, a dynamic measure of risk of the portfolio; therefore, $\pi_t = A_r(r_t, \la_t, t)/F_r(r_t, t)$.  Under this assignment, the drift and local variance become, respectively,

$$\lim_{h \rightarrow 0} {1 \over h} ({\bf E} (\Pi_{t+h} | {\cal F}_t) - \Pi) = - {\cal D}^{b^Q} A(r, \la, t) + r \Pi,  \eqno(2.13)$$

\noindent and

$$\lim_{h \rightarrow 0} {1 \over h} {\bf Var}(\Pi_{t+h} | {\cal F}_t) = \s^2(t) (\la - \ul)^2 A_\la^2(r, \la, t) + \la (A(r, \la, t) - 1)^2, \eqno(2.14)$$

\noindent in which ${\cal D}^{b^Q}$ is given by the expression in (2.9) with $b$ replaced by $b^Q = b - qd$.

Now, we come to pricing via the instantaneous Sharpe ratio.  Because the minimum local variance  in (2.14) is {\it positive}, the insurer is unable to completely hedge the mortality risk underlying the life insurance.  Therefore, the price should reimburse the insurer for its risk, say, by a constant multiple $\a$ of the local standard deviation of the portfolio, a measure of the mortality risk inherent in the insurance.  It is this $\a$ that is the instantaneous Sharpe ratio.  We could allow $\a$ to be a function of say $\la$ and $t$, to parallel the market price of risk process $\{ q_t \}$ in the bond market.  However, we choose $\a$ as a constant for simplicity.  Also, see Milevsky, Promislow, and Young (2007) for further discussion of the instantaneous Sharpe ratio.

To determine the value (price) $A$ of the life insurance, we set the drift of the portfolio equal to the short rate times the portfolio {\it plus} $\a$ times the local standard deviation of the portfolio.  Thus, from (2.13) and (2.14), we have that $A$ solves the equation

$$- {\cal D}^{b^Q} A + r \, \Pi = r \, \Pi + \a \sqrt{\s^2(t)(\la - \ul)^2 A_\la^2 + \la (A - 1)^2}, \eqno(2.15)$$

\noindent for some $0 \le \a \le \sqrt{\ul}$.  (In the proof of Theorem 3.4 the reason for the upper bound on $\a$ will become apparent.)  It follows that $A = A(r, \la, t)$ solves the non-linear pde given by

$$\left\{ \eqalign{&A_t + b^Q(r, t) A_r + {1 \over 2} d^2(r, t) A_{rr} + \m(\l, t) (\la - \ul) A_\la + {1 \over 2} \s^2(t) (\la - \ul)^2 A_{\la \la} - rA \cr
& \quad - \la (A - 1) = - \a \sqrt{\s^2(t) (\la - \ul)^2 A_\la^2 + \la (A - 1)^2}, \cr
& A(r, \la, T) = 0.} \right. \eqno(2.16)$$

\noindent Note that if the individual is still alive at time $T$, then the policy is worthless.  Hence, we have the terminal condition $A(r, \la, T) = 0$.

If we had been able to choose the investment strategy $\{ \pi_t \}$ so that the local variance in (2.14) were identically zero (that is, if the risk were hedgeable), then the right-hand side of the pde in (2.16) would be zero, and we would have a linear differential equation of the Black-Scholes type.

\subsubsect{2.2.3 Special cases for $n = 1$}

If there were no risk loading, that is, if $\a = 0$, then the price is such that the expected return on the price at time $t$ is $r_t + \la_t$.  The rate $r_t$ arises from the riskless money market, and $\la_t$ arises from the expected release of reserves as individuals die.  If $\a > 0$, then the expected return on the price is greater than $r_t + \la_t$.  Therefore, $\a$, the Sharpe ratio, measures the degree to which the insurer's total expected return is in excess of $r_t + \la_t$, as a proportion of the standard deviation of the return.  One can think of the right-hand side of (2.15) as adding a margin to the return of the portfolio because the insurance contract is not completely hedgeable due to the mortality risk.

Consider the special case for which $\s \equiv 0$, that is, the hazard rate is deterministic.  Suppose $\l(u)$, for $u \ge t$, is the solution of $d \la = \m(\la, u) (\la - \ul) du$ with initial value $\la(t) = \la$; then, under the assumption that $A \ge 0$ (which we later demonstrate in Theorem 3.4), (2.16) becomes a linear pde with solution

$$A(r, \la, t) =  \int_t^T F(r, t; s) \, e^{ - \int_t^s \left( \l(u) + \a \sqrt{\l(u)} \, \right) du} \left(\l(s) + \a \sqrt{\l(s)} \right) \, ds, \eqno(2.17)$$

\noindent in which $F(r, t; s)$ is the time-$t$ price of a bond that pays \$1 at time $s \ge t$.  The price $A$ in (2.17) is similar to the pricing rules encountered in courses in Life Contingencies because one can think of $\l(s) + \a \sqrt{\l(s)} > 0$ as a hazard rate.  Note that $A$ in (2.17) increases with respect to $\a$, as occurs more generally.

Below, we show that $A$ increases with $\a$ when the hazard rate is stochastic; see Theorem 3.7.  For this reason, we refer to the solution $A$ of (2.16) as the risk-adjusted price for a life insurance contract, in which $\a$ controls the risk adjustment.

\subsubsect{2.2.4 General case of $n \ge 1$}

To end this section, we present the pde for the price $\an$ of $n$ conditionally independent and identically distributed insurance risks.  Specifically, we assume that all the individuals are of the same age and are subject to the same hazard rate as given in (2.1); however, given that hazard rate, the occurrences of death are independent.  As discussed in the paragraph preceding equation (2.9), when an individual dies, the portfolio value $\Pi$ decreases by $1 - \left( \an - A^{(n-1)} \right)$, or increases by $\an - A^{(n-1)} - 1$.  By paralleling the derivation of (2.16), one can show that $\an$ solves the non-linear pde given by

$$\left\{ \eqalign{&\an_t + b^Q(r, t) \an_r + {1 \over 2} d^2(r, t) \an_{rr} + \m(\l, t) (\la - \ul) \an_\la + {1 \over 2} \s^2(t) (\la - \ul)^2 \an_{\la \la} \cr
& \qquad - r \an - n \la \left(\an - A^{(n-1)}  - 1 \right) \cr
& \quad = - \a \sqrt{\s^2(t) (\la - \ul)^2 \left(\an_\la \right)^2 + n \la \left(\an - A^{(n-1)} - 1 \right)^2}, \cr
&  \an(r, \la, T) = 0.} \right. \eqno(2.18)$$

\noindent The initial value in this recursion is $A^{(0)} \equiv 0$, and the price $A$ defined by (2.16) is $A^{(1)}$.

\sect{3. Qualitative Properties of $\an$}

To demonstrate a variety of properties of the risk-adjusted price $\an$, we need a comparison theorem.  We begin by stating a relevant one-sided Lipschitz condition along with growth conditions.  We require that the function $g = g(r, \la, t, v, p, q)$ satisfies the following one-sided Lipschitz condition:  For $v > w$,

$$g(r, \la, t, v, p, q) - g(r, \la, t, w, p', q') \le c_1(r, \la, t)(v - w) + c_2(r, \la, t) |p - p'| + c_3(r, \la, t) |q - q'|, \eqno(3.1)$$

\noindent with growth conditions on $c_1$, $c_2$, and $c_3$ given by

$$0 \le c_1(r, \la, t) \le K(1 + (\ln r)^2 + (\ln(\la - \ul))^2), \eqno(3.2a)$$

$$0 \le c_2(r, \la, t) \le K r (1 + |\ln r| + |\ln(\la - \ul)|), \eqno(3.2b)$$

\noindent and

$$0 \le c_3(r, \la, t) \le K (\la - \ul) (1 + |\ln r| + |\ln(\la - \ul)|), \eqno(3.2c)$$

\noindent for some constant $K \ge 0$, and for all $(r, \la, t) \in {\bf R}^+ \times (\ul, \infty) \times [0, T]$.

Throughout this paper, we rely on the following useful comparison principle, which we obtain from Walter (1970, Section 28).  We omit the proof because it is a standard application of Walter's work.

\th{3.1} {Let $G = {\bf R}^+ \times (\ul, \infty) \times [0, T],$ and denote by $\cal G$ the collection of functions on $G$ that are twice-differentiable in their first two variables, namely $r$ and $\la,$ and once-differentiable in their third, namely $t$. Define a differential operator $\cal L$ on $\cal G$ by
$${\cal L} v = v_t + {1 \over 2} d^2(r, t) v_{rr} + {1 \over 2} \s^2(t) (\la - \ul)^2 v_{\la \la} + g(r, \la, t, v, v_r, v_\la),  \eqno(3.3)$$
\noindent in which $g$ satisfies $(3.1)$ and $(3.2)$.  Suppose $v, w \in \cal G$ are such that there exists a constant $K \ge 0$ with $v \le e^{K \{(\ln r)^2 + (\ln(\la - \ul))^2\}}$ and $w \ge - e^{K \{(\ln r)^2 + (\ln(\la - \ul))^2\}}$ for large $(\ln r)^2 +(\ln(\la - \ul))^2$.  Also, suppose that there exists a constant $K \ge 0$ such that $d(r, t) \le K r$ for all $r > 0$ and $0 \le t \le T$.  \hfill \break \indent Then, if $($a$)$ ${\cal L} v \ge {\cal L} w$ on $G,$ and if $($b$)$ $v(r, \la, T) \le w(r, \la, T)$ for all $r > 0$ and $\la > \ul$, then $v \le w$ on $G$.}

\medskip

As a lemma for results to follow, we show that the differential operator associated with our problem satisfies the hypothesis of Theorem 3.1.

\lem{3.2} {Define $g_n,$ for $n \ge 1,$ by
$$\eqalign{g_n(r, \la, t, v, p, q) &= b^Q(r, t) p + \m(\la, t) (\la - \ul) q - r v  - n \la \left(v - A^{(n-1)} - 1 \right) \cr
& \quad + \a \sqrt{\s^2(t) (\la - \ul)^2 q^2 + n \la \left(v - A^{(n-1)} - 1 \right)^2},} \eqno(3.4)$$
\noindent in which $A^{(n-1)}$ solves $(2.18)$ with $n$ replaced by $n-1$.  Then, $g_n$ satisfies the one-sided Lipschitz condition $(3.1)$ on $G$.  Furthermore, if $|b^Q(r, t) | \le K r (1 + |\ln r|)$ and $|\m(\la, t)| \le K (1 + |\ln(\la - \ul)|),$ then $(3.2)$ holds.}

\pf Suppose $v > w$,  then

$$\eqalign{&g_n(r, \la, t, v, p, q) - g_n(r, \la, t, w, p', q') \cr
& \quad = b^Q(r, t) (p - p') + \m(\la, t) (\la - \ul) (q - q') - (r + n \la)(v - w) \cr
& \qquad + \a \sqrt{\s^2(t) (\la - \ul)^2 q^2 + n \la \left(v - a^{(n-1)} - 1 \right)^2} \cr 
&\qquad - \a \sqrt{\s^2(t) (\la - \ul)^2 (q')^2 + n \la \left(w - a^{(n-1)} - 1 \right)^2} \cr
& \quad \le | b^Q(r, t) | |p - p'| + \left(|\mu(\la, t)| + \a \s(t) \right) (\la - \ul) |q - q'| - \left(r + n\la - \a \sqrt{n\la} \right) (v - w) \cr
& \quad \le | b^Q(r, t) | |p - p'| + \left(|\mu(\la, t)| + \a \s(t) \right) (\la - \ul) |q - q'|.} \eqno(3.5)$$

\noindent In the first inequality, we use the fact that if $A \ge B$, then $\sqrt{C^2 + A^2} - \sqrt{C^2 + B^2} \le A - B$.  For the second inequality, recall that $0 \le \a \le \sqrt{\ul}$.  Thus, (3.1) holds with $c_1(r, \la, t) = 0$, $c_2(r, \la, t) = | b^Q(r, t) |$, and $c_3(r, \la, t) =  |\mu(\la, t)| + \a \s(t)$.  Note that $c_2$ satisfies $(3.2b)$ if $|b^Q(r, t) | \le K r (1 + |\ln r|)$, and $c_3$ satisfies $(3.2c)$ if $|\m(\la, t)| \le K (1 + |\ln(\la - \ul)|)$.  $\square$

\medskip

\noindent{\bf Assumption 3.3.} Henceforth, we assume that the volatility on the short rate $d$ satisfies the growth condition in the hypothesis of Theorem 3.1 and that the drifts $b^Q$ and $\m$ satisfy the growth conditions in the hypothesis of Lemma 3.2.  For later purposes (for example, see Theorem 3.6), we also assume that $\m_\la$ is H\"older continuous and satisfies the growth condition $|\m_\la| (\l - \ul) + | \m| \le K \left( 1 + \left( \ln (\l - \ul) \right)^2  \right)$.

\medskip

For the remainder of Section 3, we study properties of the risk-adjusted price $\an$ for $n$ life insurance contracts.  In Section 3.1, we demonstrate two basic properties of $\an$, namely, $0 \le \an \le n$ and $\an_\l \ge 0$.  In Section 3.2, we examine how $\an$ changes as we change the model parameters related to the mortality risk, specifically, $\a$, $\mu$, and $\s$.  Finally, in Section 3.3, we show that $\an$ is subadditive with respect to $n$.

\subsect{3.1  Basic Properties of $\an$}

In the first application of Theorem 3.1 and Lemma 3.2, we show that $0 \le \an \le n$ for $n \ge 0$.  Because the payoff under life insurance is nonnegative, we expect its price to be nonnegative.  Also, if the hazard rate were arbitrarily large, then we would expect the $n$ individuals to die immediately, with the corresponding price for the insurance equal to $n$.  In work in Section 4, we sharpen these bounds considerably.  Throughout, we encourage the casual reader to read the proofs lightly and to focus on the (and sometimes surprising) results of the theorems, as well as the corresponding discussions of those results.

\th{3.4} {$0 \le \an \le n$ on $G,$ for $n \ge 0$.}

\pf  We proceed by induction to prove these bounds.  The inequality holds with equality when $n = 0$ because $A^{(0)} \equiv 0$.  For $n \ge 1$, assume that $0 \le A^{(n-1)} \le (n-1)$, and show that $0 \le \an \le n$.

Define the differential operator $\cal L$ on $\cal G$ by (3.3) with $g = g_n$ from (3.4).  Because $\an$ solves (2.18), we have ${\cal L} \an = 0$.  Denote by $\bf 0$ the function that is identically $0$ on $G$; then,

$${\cal L} {\bf 0} = \left(n \l + \a \sqrt{n \l} \right) \left( A^{(n-1)} + 1 \right) \ge 0 = {\cal L} \an. \eqno(3.6)$$

\noindent Also, $\an(r, \l, T) = 0$; thus, Theorem 3.1 and Lemma 3.2 imply that $0 \le \an$ on $G$.

Next, we prove the upper bound.  Denote by $\bf n$ the function that is identically $n$ on $G$; then,

$${\cal L} {\bf n} = - \left(n \l - \a \sqrt{n \l} \right) \left( (n-1) - A^{(n-1)} \right) \le 0 = {\cal L} \an.  \eqno(3.7)$$

\noindent  Additionally, $\an(r, \la, T) = 0 < n$; thus, Theorem 3.1 and Lemma 3.2 imply that $\an \le n$ on $G$.   $\square$  

\medskip 

We end this subsection by showing that $\an_\la \ge 0$.  This result is intuitively pleasing because we expect the price of life insurance to increase as the hazard rate increases, that is, as individuals are more likely to die.  We first present a lemma which we use in the proof of Theorem 3.6.

\lem{3.5} {$\an \le A^{(n-1)} + 1$ on $G,$ for $n \ge 1.$}

\pf  We proceed by induction.  This inequality is true for $n = 1$ because $A^{(1)} \le 1 = A^{(0)} + 1$ by Theorem 3.4.  For $n \ge 2$, assume that $A^{(n-1)} \le A^{(n-2)} + 1$, and show that $\an \le A^{(n-1)} + 1$.

Define a differential operator $\cal L$ on $\cal G$ by (3.3) with $g = g_n$ from (3.4).  Because $\an$ solves (2.18), we have ${\cal L} \an = 0$.  Also,

$$\eqalign{{\cal L} \left( A^{(n-1)} + 1 \right) &= -r + (n-1) \la \left(A^{(n-1)} - A^{(n-2)} - 1 \right) + \a \s (\l - \ul)  \left| A^{(n-1)}_\l \right| \cr
& \quad - \a \sqrt{\s^2 (\la -\ul)^2 \left(A^{(n-1)}_\la \right)^2 + (n - 1) \la \left(A^{(n-1)} - A^{(n-2)} - 1 \right)^2} \cr
& \le  0 = {\cal L} \an .} \eqno(3.8)$$

\noindent  In addition, $\an(r, \la, T) = 0 < 1 = A^{(n-1)}(r, \la, T) +  1$.  Thus, Theorem 3.1 and Lemma 3.2 imply that $\an \ge A^{(n-1)} + 1$ on $G$.  $\square$

\medskip

Lemma 3.5 is interesting in its own right because it confirms our intuition that the price of adding an additional policyholder, namely $\an - A^{(n-1)}$, is less than the cost of paying the life insurance benefit immediately, namely 1.

\th{3.6} {$\an_\la \ge 0$ on $G,$ for $n \ge 0$.}

\pf  We proceed by induction.  We know that the inequality holds when $n = 0$ because $A^{(0)} \equiv 0$.  For $n \ge 1$, assume that $A^{(n-1)}_\la \ge 0$, and show that $\an_\la \ge 0$.  We apply a modified version of Theorem 3.1 to the special case of comparing $\an_\la$ with the zero function {\bf 0}.  First, differentiate $\an$'s equation with respect to $\la$ to get an equation for $f^{(n)} = \an_\la$.

$$\left\{ \eqalign{& f^{(n)}_t + b^Q f^{(n)}_r + {1 \over 2} d^2 f^{(n)}_{rr} + (\m_\la (\l - \ul) + \m - n \la - r)f^{(n)} + n \la f^{(n-1)} \cr
& \qquad + (\m + \s^2)(\la - \ul) f^{(n)}_\la +  {1 \over 2} \s^2 (\la - \ul)^2 f^{(n)}_{\la \la} - n \left(\an - A^{(n-1)} - 1 \right) \cr
& \quad = - \a { \s^2 (\la - \ul) \left(f^{(n)} \right)^2 + \s^2 (\la - \ul)^2 f^{(n)} f^{(n)}_\la + {1 \over 2} n \left(\an - A^{(n-1)} - 1 \right)^2  \over \sqrt{\s^2 (\la - \ul)^2 \left(f^{(n)} \right)^2 + n \la \left(\an - A^{(n-1)} - 1 \right)^2}} \cr
& \qquad - \a {n \la \left(\an - A^{(n-1)} - 1 \right) \left(f^{(n)} - f^{(n-1)} \right) \over \sqrt{\s^2 (\la - \ul)^2 \left(f^{(n)} \right)^2 + n \la \left(\an - A^{(n-1)} - 1 \right)^2}}, \cr
& f^{(n)}(r, \la, T) = 0.} \right. \eqno(3.9)$$

Define a differential operator $\cal L$ on $\cal G$ by (3.3) with $g = g_n$ given by

$$\eqalign{&g_n(\la, t, v, p, q) = b^Q p + (\m_\la (\l - \ul) + \m - n \la - r)v + n \la f^{(n-1)} + (\m + \s^2)(\la - \ul) q \cr
& - n \left(\an - A^{(n-1)} - 1 \right) + \a { \s^2 (\la - \ul) v^2 + \s^2 (\la - \ul)^2 v q + {n \over 2} \left(\an - A^{(n-1)} - 1 \right)^2 \over \sqrt{\s^2 (\la - \ul)^2 v^2 + n \la \left(\an - A^{(n-1)} - 1 \right)^2}} \cr
& + \a { n \la \left(\an - A^{(n-1)} - 1 \right) \left(v - f^{(n-1)} \right) \over \sqrt{\s^2 (\la - \ul)^2 v^2 + n \la \left(\an - A^{(n-1)} - 1 \right)^2}}.} \eqno(3.10)$$
 
\noindent From Walter (1970, Section 28, pages 213-215), we see that we only need to verify that (3.1) and (3.2) hold for $v > 0 = w = p' = q'$.

$$\eqalign{& g_n(\l, t, v, p, q) - g_n(\l, t, w, 0, 0) =  b^Q p + (\m_\la (\l - \ul) + \m - n \la - r)v + (\m + \s^2)(\la - \ul) q \cr
& \quad + \a { \s^2 (\la - \ul) v^2 + \s^2 (\la - \ul)^2 v q + n \l \left( \an - A^{(n-1)} - 1 \right) v  \over \sqrt{\s^2 (\la - \ul)^2 w^2 + n \la \left(\an - A^{(n-1)} - 1 \right)^2}} \cr
& \quad + {\a n \over 2} \left\{ {\left( \an - A^{(n-1)} - 1 \right)^2 \over \sqrt{\s^2 (\la - \ul)^2 v^2 + n \la \left(\an - A^{(n-1)} - 1 \right)^2}} - {\left( \an - A^{(n-1)} - 1 \right)^2 \over \sqrt{n \la \left(\an - A^{(n-1)} - 1 \right)^2}} \right\} \cr
& + \a n \l \left\{ {\left( A^{(n-1)} + 1 - \an \right) f^{(n-1)} \over \sqrt{\s^2 (\la - \ul)^2 v^2 + n \la \left(\an - A^{(n-1)} - 1 \right)^2}} - {\left( A^{(n-1)} + 1 - \an \right) f^{(n-1)} \over \sqrt{n \la \left(\an - A^{(n-1)} - 1 \right)^2}} \right\} \cr
& \le \left( \left|\m_\la \right| (\l - \ul) + |\m| + \a \s \right)(v - 0) + \left| b^Q \right| p + ( |\mu| + \s^2 + \a \s) (\l - \ul) |q|.}  \eqno(3.11)$$

\noindent Thus, by Assumption 3.3, $g_n$ satisfies (3.1) with $c_1 = |\m_\la| (\l - \ul) + |\m| + \a \s$, $c_2 = \left| b^Q \right|$, and $c_3 = (|\mu| + \s^2 + \a \s) (\l - \ul)$, in which $c_1$, $c_2$, and $c_3$ satisfy the growth conditions in (3.2).

Next, note that because $f^{(n)} = \an_\la$ satisfies (3.9), ${\cal L} f^{(n)} = 0$.  Also, we have ${\cal L} {\bf 0} = \left(n \la + \a \sqrt{n \la} \right) f^{(n-1)} + \left(n + \a/2 \sqrt{n/\la} \right) \left(A^{(n-1)} +1 - \an \right)$, which is greater than or equal to $0$ by the induction assumption and by Lemma 3.5.  These observations, together with $f^{(n)}(r, \la, T) = 0$, imply that $f^{(n)} = \an_\la \ge 0$ on $G$.  $\square$

\subsect{3.2  Comparative Statics for $A^{(n)}$}

In this section, we show that as we vary the model parameters, the price $A^{(n)}$ responds consistently with what we expect.  

\th{3.7} {Suppose $0 \le \a_1 \le  \a_2 \le \sqrt{\ul}$, and let $A^{(n), \a_i}$ be the solution of $(2.18)$ with $\a = \a_i,$ for $i = 1, 2$ and $n \ge 0$. Then, $A^{(n), \a_1} \le A^{(n), \a_2}$ on $G$.}

\pf  We proceed by induction.  It is clear that the inequality holds for $n= 0$ because $A^{(0), \a_i} \equiv 0$ for $i = 1, 2$.  For $n \ge 1$, assume that $A^{(n-1), \a_1} \le A^{(n-1), \a_2}$, and show that $A^{(n), \a_1} \le A^{(n), \a_2}$.

Define a differential operator $\cal L$ on $\cal G$ by (3.3) with $g = g_n$ from (3.4) with $\a = \a_1$.  Because $A^{(n), \a_1}$ solves (2.18) with $\a = \a_1$, we have ${\cal L} A^{(n), \a_1} = 0$.  Also,

$$\eqalign{{\cal L} A^{(n), \a_2} &= n \la \left(A^{(n-1), \a_1} - A^{(n-1), \a_2} \right) \cr
& \quad + \a_1 \left\{ \sqrt{\s^2(\la - \ul)^2 \left(A^{(n), \a_2}_\la \right)^2 + n \la \left(A^{(n), \a_2} - A^{(n-1), \a_1} - 1 \right)^2} \right.  \cr
& \qquad \qquad \left. - \sqrt{\s^2(\la - \ul)^2 \left(A^{(n), \a_2}_\la \right)^2 + n \la \left(A^{(n), \a_2} - A^{(n-1), \a_2} - 1 \right)^2} \right\} \cr
& \quad +(\a_1 - \a_2) \sqrt{\s^2(\la - \ul)^2 \left(A^{(n), \a_2}_\la \right)^2 + n \la \left(A^{(n), \a_2} - A^{(n-1), \a_2} - 1 \right)^2}  \cr
& \le \left(n \la - \a_1 \sqrt{n \la} \right) \left(A^{(n-1), \a_1} - A^{(n-1), \a_2} \right) \le 0 = {\cal L} A^{(n), \a_1}.} \eqno(3.12)$$

\noindent  In addition, both $A^{(n), \a_1}$ and $A^{(n), \a_2}$ satisfy the terminal condition $A^{(n), \a_i}(\la, T) = 0$.  Thus, Theorem 3.1 and Lemma 3.2 imply that $A^{(n), \a_1} \le A^{(n), \a_2}$ on $G$.  $\square$

\medskip

Theorem 3.7 states that as the parameter $\a$ increases, the risk-adjusted price $A^{(n), \a}$ increases.  This result justifies the use of the phrase {\it risk parameter} when referring to $\a$.  We have the following corollary of Theorem 3.7.

\cor{3.8} {Let $A^{(n), \a0}$ be the solution of $(2.18)$ with $\a = 0;$ then, $A^{(n), \a0} \le A^{(n), \a}$ for all $0 \le \a \le \sqrt{\ul},$ and we can express the lower bound $A^{(n), \a0}$ as follows:  $A^{(n), \a0} = n A^{\a0},$ in which $A^{\a0}$ is given by
$$A^{\a0} (r, \la, t) = \int_t^T F(r, t; s) \, {\bf E}^{\l, t}  \left[  e^{-\int_t^s \l_u du} \l_s \right] ds. \eqno(3.13)$$}

\pf It is straightforward to show that $n A^{\a0}$ solves (2.18) with $\a = 0$, and the result follows.  $\square$

\medskip

Note that $A^{\a0}$ is what actuaries call the {\it net premium} under the risk-neutral measure for  stochastic interest and the physical measure for stochastic mortality.  Therefore, the lower bound of ${1 \over n} A^{(n), \a}$ (as $\a$ approaches zero) is the same as the lower bound of $A^\a = A^{(1), \a}$, namely, $A^{\a0}$.  We call ${1 \over n} A^{(n)} - A^{\a0}$ the {\it risk charge} per person.  At the end of Section 4, we show how to decompose this risk charge into the charges for finite portfolio and stochastic mortality risks; see equation (4.18).

Next, we examine how the risk-adjusted price $A^{(n)}$ varies with the drift and volatility of the stochastic hazard rate.  We state the following two theorems without proof because their proofs are similar to earlier ones.

\th{3.9} {Suppose $\m_1 \le \m_2$ on $G$, and let $A^{(n), \m_i}$ denote the solution of $(2.18)$ with $\m = \m_i,$  for $i = 1, 2$ and $n \ge 0$.  Then, $A^{(n), \m_1} \le A^{(n), \m_2}$ on ${\bf R}^+ \times G$.}

From Theorem 3.9, we learn that as the {\it drift} of the hazard rate increases, then the price of life insurance increases.  This occurs for essentially the same reason that the price increases with the hazard rate; see Theorem 3.6.

\th{3.10} {Suppose $0 \le \s_1(t) \le \s_2(t)$ on $[0, T],$ and let $A^{(n), \s_i}$ denote the solution of $(2.18)$ with $\s = \s_i,$  for $i = 1, 2$ and $n \ge 0$.  If $A^{(n), \s_1}_{\la \la} \ge 0$ for all $n \ge 0,$ or if $A^{(n), \s_2}_{\la \la} \ge 0$ for all $n \ge 0,$ then $A^{(n), \s_1} \le A^{(n), \s_2}$ on ${\bf R}^+ \times G$.}

From Theorem 3.10, we see that if $\an$ is convex with respect to $\la$, then the risk-adjusted price increases as the volatility of the stochastic hazard rate increases.

\subsect{3.3  Subadditivity of $A^{(n)}$}

We next show that $A^{(n)}$ is subadditive.  Specifically, we show that for $m, n$ nonnegative integers, the following inequality holds:

$$A^{(m+n)} \le A^{(m)} + A^{(n)}.  \eqno(3.14)$$

\noindent Subadditivity is a reasonable property because if it did not hold, then buyers of insurance could  insure risks separately and thereby save money.  First, we show that the price per risk decreases with $n$, then we obtain subadditivity as a corollary.

\th{3.11} {${1 \over n} \an$ decreases with respect to $n \ge 1$ on $G$.}

\pf  We proceed by induction to show that $\an \le {n \over n-1} A^{(n-1)}$ for $n \ge 2$.  For $n = 2$, define a differential operator $\cal D$ on $\cal G$ by (3.3) with $g = g_2$ from (3.4).  Because $A^{(2)}$ solves (2.18) with $n = 2$, we have ${\cal D} A^{(2)} = 0$.  Also,

$$\eqalign{{\cal D} \left( 2 A^{(1)} \right) &= 2 \a \left\{ \sqrt{\s^2 (\la - \ul)^2 \left(A^{(1)}_\la \right)^2 + 2 \la \left({1 \over 2} A^{(1)} - {1 \over 2} \right)^2} \right. \cr
& \qquad \qquad \qquad \left. - \sqrt{\s^2 (\la - \ul)^2 \left(A^{(1)}_\la \right)^2 + (1) \la \left(A^{(1)} - A^{(0)} - 1 \right)^2} \right\} \cr
& \le 0,}  \eqno(3.15)$$

\noindent and $A^{(2)}(r, \la, T) = 0 = 2 A^{(1)}(r, \la, T)$.  Thus, it follows from Theorem 3.1 and Lemma 3.2 that $A^{(2)} \le 2 A^{(1)}$ on $G$.

Now, suppose for some $n \ge 3$, $A^{(n-1)} \le {n-1 \over n-2} A^{(n-2)}$, and show that $\an \le {n \over n-1} A^{(n-1)}$.  Define a differential operator $\cal L$ on $\cal G$ by (3.3) with $g = g_n$ from (3.4). Because $\an$ solves (2.18), we have ${\cal L} \an = 0$.  Also,

$$\eqalign{&{\cal L} \left( {n \over n-1} A^{(n-1)} \right) = n \la \left({n-2 \over n-1} A^{(n-1)} - A^{(n-2)} \right) \cr
& \quad + \a {n \over n-1} \left\{ \sqrt{\s^2 (\la - \ul)^2 \left(A^{(n-1)}_\la \right)^2 + n \la \left({1 \over n} A^{(n-1)} - {n-1 \over n} \right)^2} \right. \cr
& \qquad \qquad \qquad \left. - \sqrt{\s^2 (\la - \ul)^2 \left(A^{(n-1)}_\la \right)^2 + (n-1) \la \left(A^{(n-1)} - A^{(n-2)} - 1 \right)^2} \right\} \cr
& \le n \left(\la - \a \sqrt{\la} \right) \left({n-2 \over n-1} A^{(n-1)} - A^{(n-2)} \right) \le 0.}  \eqno(3.16)$$

The first inequality in (3.16) is not obvious, so we prove it now.  Let $B$ denote \hfill \break $\s (\l - \ul) A^{(n-1)}_\la$; then, the inequality is equivalent to

$$\eqalign{\sqrt{B^2 + n \l \left({1 \over n} A^{(n-1)} - {n-1 \over n} \right)^2} & \le \sqrt{\l} \left( (n-1) A^{(n-2)} - (n-2) A^{(n-1)} \right) \cr
& \quad + \sqrt{B^2 + (n - 1) \l \left( A^{(n-1)} - A^{(n-2)} - 1 \right)^2}.}  \eqno(3.17)$$

\noindent Square both sides of (3.17), cancel $B^2$ from both sides, and note that the resulting inequality holds for all $B$ if and only if the following holds

$$\eqalign{& {1 \over n} \left( (n-1) -  A^{(n-1)} \right)^2 \cr & \quad \le \left[ \left( (n-1) A^{(n-2)} - (n-2) A^{(n-1)} \right) + \sqrt{n-1} \left( A^{(n-2)} + 1 - A^{(n-1)} \right) \right]^2.}  \eqno(3.18)$$

\noindent Take the square root of both sides and simplify to obtain the equivalent inequality

$$0 \le A^{(n-2)} \left( (n-1) + \sqrt{n-1} \right) + A^{(n-1)} \left( -(n-2) - \sqrt{n-1} + {1 \over \sqrt{n}} \right) + \sqrt{n-1} - {n-1 \over \sqrt{n}}.    \eqno(3.19)$$

\noindent By the induction assumption, we have that $A^{(n-1)} \le {n-1 \over n-2} A^{(n-2)}$.  Thus, (3.19) holds if the following inequality holds:

$$\eqalign{0 &\le {n-2 \over n-1} A^{(n-1)} \left( (n-1) + \sqrt{n-1} \right) + A^{(n-1)} \left( -(n-2) - \sqrt{n-1} + {1 \over \sqrt{n}} \right) \cr
& \quad + \sqrt{n-1} - {n-1 \over \sqrt{n}},}    \eqno(3.20)$$

\noindent which simplifies to

$$0 \le \left[ \sqrt{n-1} - {n-1 \over \sqrt{n}} \right]  \left(1 - {1 \over n-1} A^{(n-1)}  \right).  \eqno(3.21)$$

\noindent By Theorem 3.4, we know that (3.21) holds; thus, we have shown the first inequality in (3.16).

Because ${\cal L} \left( {n \over n-1} A^{(n-1)} \right) \le 0 = {\cal L} \an$ and $\an(r, \la, T) = 0 = {n \over n-1} A^{(n-1)}(r, \la, T)$, it follows from Theorem 3.1 and Lemma 3.2 that $\an \le {n \over n-1} A^{(n-1)}$ on $G$.  $\square$

\medskip

\cor{3.12} {If $m$ and $n$ are nonnegative integers, then $A^{(m+n)} \le A^{(m)} + A^{(n)}$ on $G$.}

\pf By Theorem 3.11, ${m \over m+n} A^{(m+n)} \le A^{(m)}$ and ${n \over m+n} A^{(m+n)} \le A^{(n)}$.  Add these two inequalities to obtain inequality (3.14).   $\square$

\medskip

In the next section lies our main result.  We show that the limit (as $n$ goes to infinity) of ${1 \over n} \an$ solves a {\it linear} pde, and we give that pde.

\sect{4. Limit Result for Pricing $n$ Life Insurance Contracts}

In this section, we answer the question motivated by Theorem 3.11; that is, we determine the limiting value of the decreasing sequence ${1 \over n} A^{(n)}$ and show that the limiting value solves a linear pde, a rather surprising result.  First, we show that ${1 \over n} A^{(n)}$ is bounded below by $P$, in which $P$ solves

$$\left\{ \eqalign{&P_t + b^Q(r, t) P_r + {1 \over 2} d^2(r, t) P_{rr} + (\m(\l, t) + \a \s(t)) (\la - \ul) P_\la + {1 \over 2} \s^2(t) (\la - \ul)^2 P_{\la \la} \cr
& \quad - r P - \la \left(P - 1 \right) = 0, \cr
&  P(r, \la, T) = 0.} \right. \eqno(4.1)$$

\noindent  Intuitively, $P$ is less than ${1 \over n} \an$ because we replaced the square root in $\an$'s pde with the square root of the first term.  Later, we show that ${1 \over n} \an$ equals $P$ as $n$ approaches infinity.  For this reason, we can view $\a$ as the market price of mortality risk (in the limit).

\th{4.1} {$nP \le \an$ on $G,$ for $n \ge 0$.}

\pf  We proceed by induction.  The inequality holds for $n = 0$ because $A^{(0)} \equiv 0$.  For $n \ge 1$, assume that $(n-1) P \le A^{(n-1)}$, and show that $nP \le \an$.  Define a differential operator $\cal L$ on $\cal G$ by (3.3) with $g = g_n$ given by (3.4).  Because $\an$ solves (2.18), ${\cal L} \an = 0$.  Also,

$$\eqalign{{\cal L} (n P) &= n \l \left( A^{(n-1)} - (n-1) P) \right) \cr
& \quad + n \a \left\{ \sqrt{\s^2 (\la - \ul)^2 P^2_\la + {\la \over n} \left( nP - A^{(n-1)} - 1 \right)^2} - \s (\la - \ul) P_\la \right\} \cr
& \ge 0 = {\cal L} \an.}  \eqno(4.2)$$

\noindent In addition, $nP(r, \la, T) = \an(r, \la, T) = 0$; thus, Theorem 3.1 and Lemma 3.2 imply that $nP  \le \an$ on $G$.  $\square$

\medskip

Next, we show that $\an \le \bn$, in which $\bn$ solves

$$\left\{ \eqalign{&\bn_t + b^Q(r, t) \bn_r + {1 \over 2} d^2(r, t) \bn_{rr} + (\m(\l, t) + \a \s(t)) (\la - \ul) \bn_\la  \cr
& \quad + {1 \over 2} \s^2(t) (\la - \ul)^2 \bn_{\la \la} - r \bn - \left( n \la + \a \sqrt{n \l} \right) \left( \bn - A^{(n-1)} - 1 \right) = 0, \cr
&  \bn(r, \la, T) = 0,} \right. \eqno(4.3)$$

\noindent in which $B^{(0)} \equiv 0$.  Intuitively, $\an \le \bn$ because we replaced the square root in $\an$'s pde with the sum of the square roots of the two terms.  Recall that $\sqrt{a^2 + b^2} \le |a| + |b|$.

\th{4.2} {$\an \le \bn$ on $G,$ for $n \ge 0$.}

\pf  We proceed by induction.  The inequality holds for $n = 0$ because both $A^{(0)}$ and $\equiv B^{(0)}$ are identically zero.  For $n \ge 1$, $A^{(n-1)} \le B^{(n-1)}$, and show that $\an \le \bn$.  Define a differential operator $\cal L$ on $\cal G$ by (3.3) with $g = h_n$, in which $h_n$ is given by

$$\eqalign{h_n(r, \la, t, v, p, q) &= b^Q(r, t) p + (\m(\la, t) + \a \s(t)) (\la - \ul) q - r v \cr
& \quad - \left( n \la + \a \sqrt{n \l} \right) \left(v - A^{(n-1)} - 1 \right).} \eqno(4.4)$$

\noindent It is straightforward to show that under Assumption 3.3, $g = h_n$ in (4.4) satisfies (3.1) and (3.2).  Therefore, we can apply Theorem 3.1.  Because $\bn$ solves (4.3), ${\cal L} \bn = 0$.  Also,

$$\eqalign{{\cal L} \an & =   \a \s (\l - \ul) \an_\l + \a \sqrt{\l} \left( A^{(n-1)} + 1 - \an \right) \cr 
& \quad - \sqrt{\s^2 (\l - \ul)^2 \left( \an_\l \right)^2 +  \l \left( \an - A^{(n-1)} - 1 \right)^2} \cr
& \ge 0 = {\cal L} \bn.} \eqno(4.5)$$

\noindent Because both $\an$ and $\bn$ are zero when $t = T$, Theorem 3.1 implies that $\an \le \bn$ on $G$.  $\square$

\medskip

Finally, we come to the main result of this paper, namely, that $\lim_{n \rightarrow \infty} {1 \over n} \an = P$.

\th{4.3} {$$\lim_{n \rightarrow \infty}{1 \over n} \an = P \hbox{ on } G, \eqno(4.6)$$
in which $P$ solves $(4.1)$.}

\pf  By Theorems 4.1 and 4.2, the theorem is proved if we show that ${1 \over n} \bn - P$ goes to zero as $n$ goes to infinity because ${1 \over n} \bn - P  \ge {1 \over n} \bn - {1 \over n} \an \ge 0$.

Define $\P^{(n)}$ on $G$ by $\P^{(n)} = {1 \over n} \bn - P$, so the theorem is proved if we show that $\lim_{n \rightarrow \infty} \P^{(n)} = 0$ on $G$.  The function $\P^{(n)}$ solves the pde

$$\left\{ \eqalign{& \Pn_t + b^Q(r, t) \Pn_r + {1 \over 2} d^2(r, t) \Pn_{rr} + (\m(\l, t) + \a \s(t)) (\la - \ul) \Pn_\la  \cr
& \qquad + {1 \over 2} \s^2(t) (\la - \ul)^2 \Pn_{\la \la} - r \Pn - \left( n \la + \a \sqrt{n \l} \right) \Pn \cr
& \quad = \left( (n-1) \l +  \a \sqrt{n \l} \right) P - \left(\la + \a \sqrt{{\la \over n}} \right) A^{(n-1)} - \a \sqrt{{\l \over n}}, \cr
&\Pn(\la, T) = 0,} \right. \eqno(4.7)$$

\noindent with $0 \le \P^{(1)} = B^{(1)} - P \le 1$ because $\bn \le n$ similarly to Theorem 3.4.

We next show that if $\P^{(n-1)} \le K_{n-1}$ with $K_{n-1}$ a nonnegative constant, then $\P^{(n)} \le K_n$ via a comparison argument with a differential operator based on the pde in (4.7), in which

$$K_n = {1 \over n^{3/2}} + {n - 1 \over n} K_{n - 1}. \eqno(4.8)$$

\noindent Set $K_1 = 1$ because we know that $\P^{(1)} \le 1$.  To that end, define $\cal L$ on $\cal G$ by (3.3) with $g$ given by

$$\eqalign{g(r, \l, t, v, p, q) &= b^Q(r, t) p + (\m(\l, t) + \a \s(t))(\la - \ul) q - rv - \left(n \la + \a \sqrt{n \la} \right) v \cr
& \quad  - \left( (n-1) \l +  \a \sqrt{n \l} \right) P + \left(\la + \a \sqrt{{\la \over n}} \right) A^{(n-1)} + \a \sqrt{{\l \over n}}.}  \eqno(4.9)$$

\noindent 

\noindent It is straightforward to show that under Assumption 3.3, $g$ in (4.9) satisfies (3.1) and (3.2).  Therefore, we can apply Theorem 3.1.  Because $\Pn$ solves (4.7), we have ${\cal L} \Pn = 0$.  Assume that $\P^{(n-1)} \le K_{n-1}$ and show that $\Pn \le K_n$.  Denote the function that is identically $K_n$ in (4.8) by $\bf K_n$; then,

$$\eqalign{{\cal L} {\bf K_n} &= -r K_n - \left(n \la + \a \sqrt{n \la} \right) K_n - \left( (n-1) \l +  \a \sqrt{n \l} \right) P + \left(\la + \a \sqrt{{\la \over n}} \right) A^{(n-1)} \cr
& \quad + \a \sqrt{{\l \over n}} \cr
& = -r K_n - \left(n \la + \a \sqrt{n \la} \right) K_n +  {n-1 \over n} \left(n \la + \a \sqrt{n \la} \right) \left( {1 \over n-1} A^{(n-1)} - P \right) \cr
& \quad + \a \sqrt{{ \la \over n}} \, (1 - P).} \eqno(4.10)$$ 

\noindent Because ${1 \over n-1} A^{(n-1)} - P \le {1 \over n-1} B^{(n-1)} - P = \P^{(n-1)} \le K_{n-1}$, it follows from (4.8) and (4.10) that

$$\eqalign{{\cal L} {\bf K_n} &\le - r K_n  - {1 \over n^{3/2}} \left(n \la + \a \sqrt{n \la} \right) + \a \sqrt{{ \la \over n}} \; (1 - P) \cr
& \le -r K_n - {1 \over n^{3/2}} \left(n \la + \a \sqrt{n \la} \right) + \a \sqrt{{ \la \over n}} \le 0 = {\cal } \Pn,} \eqno(4.11)$$ 

\noindent in which the second inequality follows from $0 \le P$ and the third from $\a \le \sqrt{\ul} < \sqrt{\l}$ on $G$.  We also have $\Pn(r, \la, T) = 0 \le K_n$; thus, from Theorem 3.1, we conclude that $\Pn \le K_n$ on $G$.

Define $L_n = n K_n$.  Recall that $K_1 = 1$; thus,

$$L_n = L_{n - 1} + {1 \over \sqrt{n}}, \quad n \ge 2, \eqno(4.12)$$

\noindent from which it follows that

$$L_n = 1 + \sum_{i = 2}^n {1 \over \sqrt{i}} \le 1 + \int_1^n {dx \over \sqrt{x}} \le 1 + 2 \sqrt{n}, \quad n \ge 2. \eqno(4.13)$$

\noindent Finally, we have

$$\P^{(n)} \le K_n \le {1 \over n} + {2 \over \sqrt{n}}, \quad n \ge 1, \eqno(4.14)$$

\noindent The right-hand side of inequality (4.14) goes to zero as $n$ goes to $\infty$; thus, $\P^{(n)}$ goes to zero as $n$ goes to $\infty$.  $\square$

\medskip

We have the following corollaries of Theorem 4.3.  If the hazard rate is deterministic, then in the limit, there is no mortality risk.  The first corollary demonstrates that the price in our setting observes this desirable property and converges to the net premium. 

\cor{4.4} {If $\s \equiv 0,$ then $\lim_{n \rightarrow \infty} {1 \over n} \an = A^{\a0}$ on $G,$ in which $A^{\a0}$ is the net premium given in $(3.13)$.}

\pf From Theorem 4.3, we know that ${1 \over n} \an$ goes to $P$ as $n$ goes to infinity.  Therefore, the corollary follows because $P = A^{\a0}$ when $\s \equiv 0$, which is clear from (4.1) by setting $\s$ equal to 0.  $\square$

\medskip

On the other hand, if the hazard rate is truly stochastic, then in the limit, mortality risk remains.  The next corollary shows that the limiting price in this case is strictly greater than the net premium when $t < T$.

\cor{4.5} {If $\s$ is uniformly bounded below by $\kappa > 0$, then $\lim_{n \rightarrow \infty} {1 \over n} A^{(n)} \ge A^{\a0}$ on $G,$ with equality only when $t = T$.}

\pf This result follows from the fact that $P \ge A^{\a0},$ with equality only when $t = T$.  Indeed, define a differential operator $\cal L$ on $\cal G$ by (3.3) with $g = g_1$ in (3.4) and with $\a = 0$.  Thus, ${\cal L} A^{\a0} = 0$, and

$${\cal L} P = - \a \s (\la - \ul) P_\la \le 0 = {\cal L} A^{\a0}.  \eqno(4.15)$$

\noindent Also, $A^{\a0}(r, \la, T) = P(r, \la, T) = 0$; thus, Theorem 3.1 and Lemma 3.2 imply that $P \ge A^{\a0}$.

To show that $P < A^{\a0}$ for $t \in [0, T)$, consider the pde of $C = P - A^{\a0}$.  The function $C$ solves

$$\left\{ \eqalign{& C_t + b^Q C_r + {1 \over 2} d^2 C_{rr} + (\m +  \a \s) (\l - \ul) C_\l + \a \s A^{\a0}_\l \cr
& \quad + {1 \over 2} \s^2 (\la - \ul)^2 C_{\l \l} - rC - \l \left(C - 1 \right) = 0, \cr
& C(r, \la, T) = 0.} \right. \eqno(4.16)$$

\noindent From the linear pde in (4.16) and the Feynman-Kac Theorem, we deduce that

$$C(r, \la, t) = \left({\bf \hat E^Q} \right)^{r, \l, t} \left[ \int_t^T \left(\l_s + \a \s(s) A^{\a0}_\l(r_s, \la_s, s) \right) e^{- \int_t^s (r_u + \la_u) du} ds  \right], \eqno(4.17)$$

\noindent in which $\{ \la_s \}$ follows the diffusion $d \la_s = (\m(\l_s, s) + \a \s(s))(\la_s - \ul) ds + \s(s)(\la_s - \ul) d \hat W^\la_s$, with $\hat W^\la_s = W^\la_s - \a s$.  Also, $\{ r_s \}$ follows the diffusion $d r_s = b^Q(r_s, s) ds + d(r_s, s) d W^Q_s$, with $W^Q_s = W_s + \int_0^s q(r_u, u) du$; see the discussion following (2.3).  Note that $\a$ is analogous to the bond market's price of risk $\{ q (r_s, s) \}$.  The processes $\{ \hat W^\la_s \}$ and $\{ W^Q_s \}$ are independent standard Brownian motions with respect to a suitably-defined probability space, and $\bf \hat E^Q$ denotes expectation on that space.

From the representation of $C = P - A^{\a0}$ in (4.17), we see that $C(r, \l, t) > 0$ for $t < T$ because $\l_s > \ul$ and $A^{\a0}_\l \ge 0$.  $\square$

\medskip

After all this work, we can finally decompose the risk charge ${1 \over n} \an - A^{\a0}$ -- first mentioned in the Introduction and again following Corollary 3.8 -- into its component risk charges:  one for a finite portfolio and another for stochastic mortality.  Recall from Theorem 4.3 that $\lim_{n \rightarrow \infty} {1 \over n} \an = P$, in which $P$ solves (4.1).  Therefore, define ${1 \over n} A^{(n)} - P$ as the risk charge (per risk) for holding a finite portfolio, and  define $P - A^{\a0}$ as the risk charge for stochastic mortality even after selling to an arbitrarily large group.  Thus, we have

$${1 \over n} A^{(n)} - A^{\a0} = \left({1 \over n} A^{(n)} - P \right) + \left(P - A^{\a0} \right), \eqno(4.18)$$

\noindent in which the risk charge for stochastic mortality, namely $P - A^{\a0}$, is zero if $\s \equiv 0$ by Corollary 4.4 and is positive (for $t < T$) if $\s \ge \kappa > 0$ by Corollary 4.5.

We end this section by more closely examining the limiting value of ${1 \over n} \an$, namely $P$, and show that it can be represented as an expectation with respect to an equivalent martingale measure.   Note the similarity between the limiting price $P$ in (4.1) and the net premium $A^{\a0}$ in (3.13), or equivalently (2.16) with  $\a = 0$.

\cor{4.6} {One can represent the limiting value of ${1 \over n} \an,$ namely $P,$ as follows:
$$P(r, \l, t) = \int_t^T  F(r, t; s) \, {\bf \hat E}^{\l, t} \left[ \l_s e^{- \int_t^s \l_u du} \right] \, ds,  \eqno(4.19)$$
in which $\bf \hat E$ is the expectation on the probability space $(\Omega, {\cal F}, {\bf \hat P}),$ in which the Radon-Nikodym derivative of $\bf \hat P$ with respect to $\bf P$ is given by
$${d {\bf \hat P} \over d {\bf P}} \bigg|_{{\cal F}_t} = \exp \left( \a W_t - {1 \over 2} \a^2 t \right).  \eqno(4.20)$$
Note that $\{ \hat W^\l_s \}$ as defined in the paragraph following $(4.17)$ is a standard Brownian motion on $(\Omega, {\cal F}, {\bf \hat P},$ and recall that $\{ \la_s \}$ has dynamics
$$d \la_s = (\m(\l_s, s) + \a \s(s))(\la_s - \ul) ds + \s(s)(\la_s - \ul) d \hat W^\la_s. \eqno(4.21)$$}

\pf From the Feynman-Kac Theorem applied to $P$'s linear pde (4.1); see Karatzas and Shreve (1991), we can represent $P$ as follows

$$\eqalign{P(r, \l, t) &= \left( {\bf \hat E^Q} \right)^{r, \l, t} \left[ \int_t^T  \l_s e^{- \int_t^s (r_u + \l_u) du} \, ds \right] \cr
&= \int_t^T  \left({\bf E^Q}\right)^{r, t} \left[ e^{- \int_t^s r_u du} \right] \, {\bf \hat E}^{\l, t} \left[ \l_s e^{- \int_t^s \l_u du} \right] \, ds \cr
&= \int_t^T  F(r, t; s) \, {\bf \hat E}^{\l, t} \left[ \l_s e^{- \int_t^s \l_u du} \right] \, ds.}  \eqno(4.22)$$

$\square$

\medskip

Note that the representation of $P$ in (4.19) is quite similar to a net single premium as defined in courses on Life Contingencies.  Indeed, $F$ is the monetary discount function, and $f(\l, t; s) = {\bf \hat E}^{\l, t} \left[ \l_s e^{- \int_t^s \l_u du} \right]$ plays the role of the probability density function of the time of death.  Also, note that we can see clearly in (4.21) that $\a$ in the insurance market is parallel to the bond market's price of risk $\{ q(r_s, s) \}$.

It is clear from Corollary 4.6, for example, that we can essentially use any (reasonable) model for bond pricing.  Additionally, one could use a different sort of default-free bond, such as a consol bond (or perpetuity).  The latter would be more appropriate in the setting of whole life insurance.  Alternatively, an insurer could continually roll money into longer-term bonds as the shorter-term bonds mature.

For later reference, note that $f$ solves the pde

$$\left\{ \eqalign{&f_t + (\m(\l, t) + \a \s(t)) (\la - \ul) f_\la + {1 \over 2} \s^2(t) (\la - \ul)^2 f_{\la \la} - \la f = 0, \cr
&  f(\la, s; s) = \l.} \right. \eqno(4.23)$$

\noindent In the next section, we examine a numerical example.

\sect{5. Numerical Example}

We want to determine how well the lower bound $P$ in (4.1) approximates the price $A$ in (2.16), or more generally ${1 \over n} \an$ in (2.18), for a specific numerical example.  Because $P$ is a lower bound, then we know that it is biased low.  From (4.14), it follows that $P \le {1 \over n} \an \le {1 \over n} \bn \le P + {1 \over n} + {2 \over \sqrt{n}}$.  Due to the slow rate of convergence of ${1 \over n} \bn$ (and presumably of ${1 \over n} \an$) to $P$ (namely, of the order $O(1/\sqrt{n})$), we also consider the upper bound $B = B^{(1)}$ that solves (4.3) with $n = 1$.  If $B$ provides a better estimate for $A$ than $P$, then it will provide a better estimate for ${1 \over n} \an$ for $n$ small enough.

The upper bound $B$ solves

$$\left\{ \eqalign{&B_t + b^Q(r, t) B_r + {1 \over 2} d^2(r, t) B_{rr} + (\m(\l, t) + \a \s(t)) (\la - \ul) B_\la  \cr
& \quad + {1 \over 2} \s^2(t) (\la - \ul)^2 B_{\la \la} - r B - \left( n \la + \a \sqrt{n \l} \right) \left( B - 1 \right) = 0, \cr
&  B(r, \la, T) = 0,} \right. \eqno(5.1)$$

\noindent Similarly to Corollary 4.6, we can represent $B$ as follows:

$$B(r, \l, t) = \int_t^T  F(r, t; s) \, {\bf \hat E}^{\l, t} \left[ \left(\l_s + \a \sqrt{\l_s} \right) e^{- \int_t^s \left(\l_u + \a \sqrt{\l_u} \right) du} \right] \, ds,  \eqno(5.2)$$

\noindent Let $g$ denote the function given by $g(\l, t; s) = {\bf \hat E}^{\l, t} \left[ \left(\l_s + \a \sqrt{\l_s} \right) e^{- \int_t^s \left(\l_u + \a \sqrt{\l_u} \right) du} \right]$; then, $g$ solves

$$\left\{ \eqalign{&g_t + (\m(\l, t) + \a \s(t)) (\la - \ul) g_\la + {1 \over 2} \s^2(t) (\la - \ul)^2 g_{\la \la} - (\la + \a \sqrt{\l}) g = 0, \cr
&  g(\la, s; s) = \l + \a \sqrt{\l}.} \right. \eqno(5.3)$$

\noindent Note the parallel between $g$ in (5.3) and $f$ in (4.23).

It follows from (4.19) and (5.2) that to compute $P$ or $B$ numerically, it is enough to compute $f$ or $g$, respectively, and then numerically integrate the result with $F$.  Because the non-linearity in the equation for $A$ arises from the stochastic hazard rate, to test the tightness of the bounds $P$ and $B$, we assume that the interest rate $r$ is identically zero and, thereby, focus on the effect of the stochastic hazard rate.  Note that in this case, the discount function $F$ is identically 1.

For this example, we assume that the hazard rate follows the diffusion in (2.1) with $\m$ and $\s$ constant.  In this case,

$$\l_t = \ul + (\l_0 - \ul) \exp \left( \left( \m - {1 \over 2} \s^2 \right) t + \s W^\l_t \right), \eqno(5.4)$$

\noindent a stochastic version of Makeham's law (Bowers et al., 1997).

See Appendix A for an algorithm for numerically computing the lower bound $P$.  The one for the upper bound $B$ is obtained similarly, and for the sake of space, we omit it.  Because we assume that $r$ is identically zero, we can also solve for the risk-adjusted prive $A$ numerically by using a similar algorithm, and this allows us to see which of $P$ or $B$ gives us the better approximation to $A$.  See Appendix B for an algorithm to compute $A$.

We use the following values of the parameters for the computation in Table 1:

\item{$\bullet$} The term of the insurance is $T = 10$ years.

\item{$\bullet$} The minimum value of the hazard rate is $\ul = 0.02$.

\item{$\bullet$} The drift of the hazard rate $\m = 0.04$.

\item{$\bullet$} The volatility of the hazard rate $\s = 0.10$.

\item{$\bullet$} The risk parameter $\a = 0.10$.

In Table 1, for a variety of initial values of the hazard rate, we present the limiting price $P$, the risk-adjusted price $A$, and the upper bound $B$.  For reference, we also include the physical probability of dying $A^{\a0}$.  Note that in each case, the upper bound $B$ provides a better approximation to the risk-adjusted price $A$ than does the limiting price $P$.  For other numerical experiments (not shown here), we found the same phenomenon, namely that $B$ is a better approximation to $A$ than $P$.

\medskip

\settabs 5 \columns \+  $\l_0$Ê &Ê$A^{\a0}$ & $P$ & $A$ & $B$ \cr \smallskip \hrule \smallskip
\+ 0.020 &  0.1813 & 0.1817 & 0.2896 & 0.2897 \cr
\+ 0.021 &  0.1914 & 0.1919 & 0.3010 & 0.3017 \cr
\+ 0.022 &  0.2014 & 0.2025 & 0.3126 & 0.3139 \cr
\+ 0.023 &  0.2112 & 0.2128 & 0.3237 & 0.3256 \cr
\+ 0.024 & 0.2214 & 0.2235 & 0.3352 & 0.3377 \cr
\+ 0.025 & 0.2300 & 0.2326 & 0.3449 & 0.3477  \cr
\+ 0.030 & 0.2763 & 0.2812 & 0.3953 & 0.4004  \cr
\+ 0.035 & 0.3187 & 0.3256 & 0.4397 & 0.4466  \cr
\+  0.040 & 0.3609 & 0.3696 & 0.4826 & 0.4909  \cr
\+  0.050 & 0.4338 & 0.4451 & 0.5536 & 0.5639  \cr
\+  0.060 & 0.5017 & 0.5150 & 0.6169 & 0.6285  \cr
\+  0.070 & 0.5530 & 0.5675 & 0.6630 & 0.6753  \cr \smallskip \hrule \medskip

\medskip

\sect{6. Summary and Conclusions}

We developed a risk-adjusted pricing method for life insurance by assuming that the insurance company is compensated for its risk in the form of a given instantaneous Sharpe ratio of an appropriately-defined (partially) hedging portfolio.  Because the market for insurance is incomplete, one cannot assert that there is a unique price.  However, we believe that the price that our method produces is a valid one because of the many desirable properties that it satisfies.  In particular, we studied properties of the price of $n$ conditionally independent and identically distributed life insurance contracts.  In Theorem 3.11, we showed that the risk charge per person decreases as $n$ increases, and in Corollary 3.12, we showed that the price is subadditive with respect to $n$.

Arguably our main results are those dealing with the limiting price per person.  Our most important result is that the limiting price solves a linear pde (Theorem 4.3), and we provided a probabilistic interpretation of the limiting price as an expectation with respect to an equivalent martingale measure (Corollary 4.6).  Our work, therefore, extends that of Blanchet-Scalliet, El Karoui, and Martellini (2005) and Dahl and M\o ller (2006).

We proved that if the hazard rate is deterministic, then the risk charge per person goes to zero as $n$ goes to infinity (Theorem 4.3 and Corollary 4.4).  Moreover, we proved that if the hazard rate is stochastic, then the risk charge person is positive as $n$ goes to infinity, which reflects the fact that the mortality risk is not diversifiable in this case (Theorem 4.3 and Corollary 4.5).  Additionally, in equation (4.18), we decomposed the per-risk risk charge into the finite portfolio and stochastic mortality risk charges.

Milevsky, Promislow, and Young (2005, 2007) proved similar properties for the risk-adjusted prices of pure endowments and annuities; see Milevsky, Promislow, and Young (2007) for an elementary discussion of this general topic.  Because of these properties, we anticipate that our pricing methodology will prove useful in pricing risks for many insurance products.  For example, Bayraktar and Young (2007a) apply the method in this paper to price both pure endowments and life insurance.  They showed that the price of the two products combined is less than the sum of the individual prices, an intuitively pleasing result.  An interesting extension would be to allow the hazard rate to exhibit jumps, as a model for catastrophes.

We also believe that our valuation method will be useful in pricing risks in other incomplete markets.  For example, Bayraktar and Young (2007b) price options on non-traded assets for which there is a traded asset that is correlated to the
non-traded asset.  Then, they to price options in the presence of stochastic volatility, and the instantaneous Sharpe ratio in this case equals the market price of volatility risk, similar to interpreting the instantaneous Sharpe ratio in this paper as the market price of mortality risk in (4.1).

\sect{Appendix A.  Algorithm for computing $P$}

We present this algorithm in some detail so that the interested reader can reproduce our results (or related results).  First, we transform the equation for $f$, namely (4.23), by defining $\tau = s - t$, $y = \ln(\l - \ul)$, and  $\tilde f(y, \tau) = f(\ul + e^y, s - \tau)$.  Then, $\tilde f$ solves

$$\left\{ \eqalign{& \tilde f_\tau = \hat \m \tilde f_y + {1 \over 2} \s^2 \tilde f_{yy} - (\ul  + e^y) \tilde f, \cr
&  \tilde f(y, 0) = \ul + e^y,} \right. \eqno({\rm A.1})$$

\noindent in which $\hat \m = \m + \a \s - {1 \over 2} \s^2$.  Instead of solving (A.1) for $(y, \tau)$ in the domain ${\bf R} \times [0, T]$, we solve (A.1) on the domain $[-M, M] \times [0, T]$, for $M$ such that $e^{-M}$ is approximately zero.  Therefore, we require boundary conditions $\tilde f(-M, \tau)$ and $\tilde f(M, \tau)$.

If $\l_t = \ul$, then $\l_s = \ul$ for all $s \ge t$.  Thus, $f(\ul, t; s) = \ul \, e^{-\ul(s - t)}$, and it follows that the appropriate boundary condition at $y = -M$ is $\tilde f(-M, \tau) = \ul \, e^{-\ul \tau}$.

If $\l_t$ is large, then we expect the individual to die immediately, so that $f(\l, s)$ is approximately 0 for $s > t$.  Thus, the boundary condition at $y = M$ is $\tilde f(M, \tau) = 0$ for $\tau > 0$.

Subdivide $[-M, M] \times [0, T]$ into a grid of rectangles of size $h$ by $k$.  Thus, define $y_n = - M + nh$ for $n = 0, 1, \dots, N$  with $N = 2M/h$, and define $\tau_j = jk$ for $j = 0, 1, \dots, J$ with $J = T/k$.  Assume that $M$ and $h$ are such that $N$ is an integer; similarly, assume that $k$ is such that $J$ is an integer.  Now, define $\tilde f_{nj} = \tilde f(y_n, \tau_j)$.

We use a backward difference (or implicit) scheme because it is inherently more stable (DuChateau and Zachmann, 1986).  Note that

$$\tilde f_\tau(y_n, \tau_j) = {\tilde f_{n, j+1} - \tilde f_{nj} \over k} + O(k), \eqno({\rm A.2})$$

$$\tilde f_y(y_n, \tau_j) = {\tilde f_{n+1, j+1} - \tilde f_{n-1, j+1} \over 2h} + O(h^2), \eqno({\rm A.3})$$

\noindent and

$$\tilde f_{yy}(y_n, \tau_j) = {\tilde f_{n+1, j+1} -2 \tilde f_{n, j+1} + \tilde f_{n-1, j+1} \over h^2} + O(h^2), \eqno({\rm A.4})$$

\noindent so that the following equation approximates (A.1) at $(y, \tau) = (y_n, \tau_j)$ up to order $O(k + h^2)$.

$$\eqalign{ {\tilde f_{n, j+1} - \tilde f_{nj} \over k} &= \hat \m {\tilde f_{n+1, j+1} - \tilde f_{n-1, j+1} \over 2h} + {1 \over 2} \s^2 {\tilde f_{n+1, j+1} -2 \tilde f_{n, j+1} + \tilde f_{n-1, j+1} \over h^2} \cr
& \quad - \left(\ul + e^{y_n} \right) \tilde f_{n, j+1}.}    \eqno({\rm A.5})$$

If we define $a = \hat \m {k \over 2h} - {1 \over 2} \s^2 {k \over  h^2}$, $b = 1 + \s^2 {k \over h^2} + k \ul$, and $c = -\hat \mu {k \over 2h} - {1 \over 2} \s^2 {k \over h^2}$, then (A.5) becomes

$$a \tilde f_{n-1, j+1} + \left( b + k e^{y_n} \right) \tilde f_{n, j+1} + c \tilde f_{n+1, j+1} = \tilde f_{nj}, \eqno({\rm A.6})$$

\noindent for $n = 2, 3, \dots, N-2$ and $j = 0, 1, \dots, J-1$, with the boundary conditions giving us

$$\left( b + e^{y_1} \right) \tilde f_{1, j+1} + c \tilde f_{2, j+1} = \tilde f_{1j} - a \ul e^{-\ul (j+1)k}, \quad \hbox{for } j = 0, 1, \dots, J-1, \eqno({\rm A.7})$$

\noindent and

$$a \tilde f_{N-1, j+1} + \left( b + k e^{y_{N-1}} \right) \tilde f_{N-1, j+1} = \tilde f_{N-1, j}, \quad \hbox{for } j = 0, 1, \dots, J-1. \eqno({\rm A.8})$$

The system of equations in (A.6) - (A.8) can be represented in matrix form by

$${\bf M} \, {\bf \tilde f}_{j+1} = {\bf \tilde f}_{j} - \left[ a \ul e^{-\ul (j+1)k}, 0, \dots, 0 \right]^t, \quad \hbox{for } j = 0, 1, \dots, J-1. \eqno({\rm A.9})$$

\noindent in which the superscript $t$ denotes matrix transform and $\bf M$ is the tri-diagonal matrix with main diagonal given by the terms $b + k e^{y_1}, b + k e^{y_2}, \dots, b + k e^{y_{N-1}}$, with the sub-diagonal identically $a$, and with the super-diagonal identically $c$.  Finally, ${\bf \tilde f}_j$ denotes the column vector $\left[ \tilde f_{1 j}, \tilde f_{2 j}, \dots, \tilde f_{N-1, j} \right]^t$, and similarly for ${\bf \tilde f}_{j+1}$.

Beginning with the initial value $\tilde f_{n 0} = \ul + e^{y_n}$ for $n = 1, 2, \dots, N-1$, we solve (A.9) repeatedly until we obtain $\tilde f_{n J}$ for $n = 1, 2, \dots, N-1$.  We use the Thomas algorithm for solving (A.9), which relies on the LU-decomposition of $\bf M$ into a product of a lower- and an upper-diagonal matrix.

Once we have $\tilde f$, then we want to estimate $P$ at time 0.

$$\eqalign{P(r, \l, 0) &= \int_0^T F(r, 0; s) f(\l, 0; s) ds \cr
& \approx \left[ {1 \over 2} F(r, 0; 0) f(\l, 0; 0) + \sum_{j=1}^{J-1} F(r, 0; jk) f(\l, 0; jk) + {1 \over 2} F(r, 0; T) f(\l, 0; T) \right] k.}  \eqno({\rm A.10})$$

\noindent Suppose that $\l = \ul + e^y = \ul + e^{y_n}$, then (A.10) becomes

$$P(r, \l, 0) \approx {1 \over 2} \left[ F(r, 0; 0) \tilde f_{n0} + \sum_{j=1}^{J-1} F(r, 0; jk) \tilde f_{nj} + {1 \over 2} F(r, 0; T) \tilde f_{nJ} \right] k,  \eqno({\rm A.11})$$

\noindent which becomes the following simplified expression because we assume that $r$ is identically zero, or equivalently $F$ is identically one:

$$P(r, \l, 0) \approx {1 \over 2} \left[ \tilde f_{n0} + \sum_{j=1}^{J-1} \tilde f_{nj} + {1 \over 2} \tilde f_{nJ} \right] k.  \eqno({\rm A.12})$$

\noindent In a sense, the expression in (A.12) is a modified probability of dying before time $T$.  It will be larger than the physical probability of dying, which is measured when $\a = 0$.  However, $P$ will be less than the corresponding value for the price $A$ and the latter less than the upper bound $B$.

In pseudo-code form, here is the detailed algorithm for computing $\tilde f_{nj}$ for $n = 1, 2, \dots, \break N-1$ and for $j = 1, 2, \dots, J$:

\item{1. } Define $M$ so that $e^{-M}$ is approximately zero, or equivalently, so that $e^M$ is quite large.

\item{2. } Define $T$ as the horizon for the life insurance policy.  To price whole life insurance, set $T$ equal to some large value, such as 300.

\item{3. } Define the $y$-step-size $h$, so that $N = 2M/h$ is an integer.

\item{4. } Define the $\tau$-step-size $k$, so that $J = T/k$ is an integer.

\item{5. } Define the parameters of the stochastic hazard diffusion $\m$, $\s$, and $\ul$, and define the pricing parameter $\a \in [0, \sqrt{\ul})$.  Then, define $\hat \m = \m + \a \s - {1 \over 2} \s^2$.

\item{6. } Define $a = \hat \m {k \over 2h} - {1 \over 2} \s^2 {k \over  h^2}$, $b = 1 + \s^2 {k \over h^2} + k \ul$, and $c = -\hat \mu {k \over 2h} - {1 \over 2} \s^2 {k \over h^2}$.

\item{7. } For $n = 1, 2, \dots, N-1$, define $y_n = -M + nh$.

\item{8. } For $n = 1, 2, \dots, N-1$, define $\tilde f_{n0} = \ul + e^{y_n}$.
 
\item{9. } Define $l_1 = b + k e^{y_1}$.  Then, for $n = 2, 3, \dots, N-1$, define $l_n = b + k e^{y_n} - ac/l_{n-1}$.

\item{10. } For $j = 0, 1, \dots, J-1$, define $z_{1j} = {1 \over l_1} \left( \tilde f_{1j} - a \ul e^{-\ul (j+1) k} \right)$, and for $n = 2, 3, \dots, N-1$, define $z_{nj} = {1 \over l_n} \left( \tilde f_{nj} - a z_{n-1, j} \right)$.  Then, still within the $j$-loop, define $\tilde f_{N-1, j+1} = z_{N-1, j}$ and for $n = N-2, N-3, \dots, 1$, define $\tilde f_{n, j+1} = z_{nj} - {c \over l_n} \tilde f_{n+1, j+1}$.  $\square$

\medskip

\noindent For the work in this paper, we set $M = 10$ ($e^{-10} = 4.54 \times 10^{-5}$, or equivalently, $e^{10}$ = 22,026), $h = 0.1$, and $k = 0.01$.

\sect{Appendix B.  Algorithm for computing $A$}

The algorithm to compute $A$ is based on the one in Appendix A, so we simply write the algorithm.  One difference between this algorithm and the one in Appendix A is that we approximate the derivative $A_\l$ in the non-linear term on the right-hand side of (2.16) with a forward difference and include that non-linear term on the right-hand side of the analog of (A.9).  Another difference lies in the boundary conditions for small and large values of $\l$: $A(0, \ul, t) = 1 - e^{- \left( \ul + \a \sqrt{\ul} \right) (T - t)}$, and $A(0, \l, t) = 1$ for $\l$ large.

In pseudo-code form, here is the detailed algorithm for computing $\tilde A_{nj}$ for $n = 1, 2, \dots, \break N-1$ and for $j = 1, 2, \dots, J$ when $r$ is identically zero:

\item{1. } Define $M$ so that $e^{-M}$ is approximately zero, or equivalently, so that $e^M$ is quite large.

\item{2. } Define $T$ as the horizon for the life insurance policy.  To price whole life insurance, set $T$ equal to some large value, such as 300.

\item{3. } Define the $y$-step-size $h$, so that $N = 2M/h$ is an integer.

\item{4. } Define the $\tau$-step-size $k$, so that $J = T/k$ is an integer.

\item{5. } Define the parameters of the stochastic hazard diffusion $\m$, $\s$, and $\ul$, and define the pricing parameter $\a \in [0, \sqrt{\ul})$.  Then, define $\m' = \m - {1 \over 2} \s^2$.

\item{6. } Define $a = \m' {k \over 2h} - {1 \over 2} \s^2 {k \over  h^2}$, $b = 1 + \s^2 {k \over h^2} + k \ul$, $c = -\mu' {k \over 2h} - {1 \over 2} \s^2 {k \over h^2}$, and $G = {1 \over 2} \s^2 \left( {k \over 2h} \right)^2$.

\item{7. } For $n = 1, 2, \dots, N-1$, define $y_n = -M + nh$.

\item{8. } For $n = 1, 2, \dots, N-1$, define $\tilde A_{n0} = 0$.
 
\item{9. } Define $l_1 = b + k e^{y_1}$.  Then, for $n = 2, 3, \dots, N-1$, define $l_n = b + k e^{y_n} - ac/l_{n-1}$.

\item{10. } For $j = 0, 1, \dots, J-1$, define

$$z_{1j} = {1 \over l_1} \left[ \tilde A_{1j} + k \left( \ul + e^{y_1} \right) - a \left( 1 - e^{- \left( \ul + \a \sqrt{\ul} \right) (j+1) k} \right) + \a S_{1j} \right],$$

\hangindent 20 pt in which $S_{1j} = \sqrt{G \left( \tilde A_{2j} - 1 + e^{-\left(\ul + \a \sqrt{\ul} \right) jk} \right)^2 + k^2 \left( \ul + e^{y_1} \right) \left( \tilde A_{1j} - 1 \right)^2}$,  and for $n = 2, 3, \dots, N-2$, define

$$z_{nj} = {1 \over l_n} \left[ \tilde A_{nj} + k \left( \ul + e^{y_n} \right) - a z_{n-1, j} + \a S_{nj} \right],$$

\hangindent 20 pt in which $S_{nj} = \sqrt{G \left( \tilde A_{n+1, j} - \tilde A_{n-1, j} \right)^2 + k^2 \left( \ul + e^{y_n} \right) \left( \tilde A_{nj} - 1 \right)^2}$.  Finally, define

$$z_{N-1, j} = {1 \over l_{N-1}} \left[ \tilde A_{N-1, j} + k \left( \ul + e^{y_{N-1}} \right) - c - a z_{N-2, j} + \a S_{N-1, j} \right],$$

\hangindent 20 pt in which $S_{nj} = \sqrt{G \left( 1 - \tilde A_{N-2, j} \right)^2 + k^2 \left( \ul + e^{y_{N-1}} \right) \left( \tilde A_{N-1, j} - 1 \right)^2}$. Then, still with-in the $j$-loop, define $\tilde A_{N-1, j+1} = z_{N-1, j}$ and for $n = N-2, N-3, \dots, 1$, define $\tilde A_{n, j+1} = z_{nj} - {c \over l_n} \tilde A_{n+1, j+1}.$  $\square$

\sect{Acknowledgements}

I thank The Actuarial Foundation for financially supporting my work, and I thank S. David Promislow, Erhan Bayraktar, and Michael Ludkovski for their valuable help.

\sect{References}

\noindent \hangindent 20 pt Ballotta, L. and S. Haberman (2006), The fair valuation problem of guaranteed annuity options: the stochastic mortality environment case, {\it Insurance: Mathematics and Economics}, 38: 195-214.

\smallskip \noindent \hangindent 20 pt Bayraktar, E. and V. R. Young (2007), Hedging life insurance with pure endowments, {\it Insurance: Mathematics and Economics}, to appear.

\smallskip \noindent \hangindent 20 pt Biffis, E. (2005), Affine processes for dynamic mortality and actuarial valuation, {\it Insurance: Mathematics and Economics}, to appear.

\smallskip \noindent \hangindent 20 pt Bj\"ork, T. (2004), {\it Arbitrage Theory in Continuous Time}, second edition, Oxford University Press, Oxford.

\smallskip \noindent \hangindent 20 pt Bj\"ork, T. and I. Slinko (2006), Towards a general theory of good deal bounds, {\it Review of Finance}, 10: 221-260.

\smallskip \noindent \hangindent 20 pt Blanchet-Scalliet, C., N. El Karoui, and L. Martellini (2005), Dynamic asset pricing theory with uncertain time-horizon, {\it Journal of Economic Dynamics and Control}, 29: 1737-1764.

\noindent \hangindent 20 pt Bowers, N. L., H. U. Gerber, J. C. Hickman, D. A. Jones, and C. J. Nesbitt (1997), {\it Actuarial Mathematics}, second edition, Society of Actuaries, Schaumburg, Illinois.

\smallskip \noindent \hangindent 20 pt Cairns, A. J. G., D. Blake and K. Dowd (2004), Pricing framework for securitization of mortality risk, working paper, Heriot-Watt University.

\smallskip \noindent \hangindent 20 pt Cox, S. H. and Y. Lin (2004), Natural hedging of life and annuity mortality risks, working paper, Robinson College of Business, Georgia State University.

\smallskip \noindent \hangindent 20 pt Dahl, M. (2004), Stochastic mortality in life insurance: market reserves and mortality-linked insurance contracts, {\it Insurance: Mathematics and Economics}, 35 (1): 113-136.

\smallskip \noindent \hangindent 20 pt Dahl, M. and T. M\o ller (2006), Valuation and hedging of life insurance liabilities with systematic mortality risk, {\it Insurance: Mathematics and Economics}, 39 (2): 193-217.

\smallskip \noindent \hangindent 20 pt DuChateau, P. and D. W. Zachmann (1986), {\it Schaum's Outline of Partial Differential Equations}, McGraw-Hill, New York.

\smallskip \noindent \hangindent 20 pt Gavrilov, L. A. and N. S. Gavrilova (1991), {\it The Biology of Life Span: A Quantitative Approach}, Harwood Academic Publishers, New York.

\smallskip \noindent \hangindent 20 pt Karatzas, I. and S. E. Shreve (1991), {\it Brownian Motion and Stochastic Calculus}, second edition, Springer-Verlag, New York.

\smallskip \noindent \hangindent 20 pt Lamberton, D. and B. Lapeyre (1996), {\it Introduction to Stochastic Calculus Applied to Finance}, Chapman \& Hall/CRC, Boca Raton, Florida.

\smallskip \noindent \hangindent 20 pt Lee, R. and L. Carter (1992), Modeling and forecasting U. S. mortality, {\it Journal of the American Statistical Association}, 87: 659-671.

\smallskip \noindent \hangindent 20 pt Milevsky, M. A. and S. D. Promislow (2001), Mortality derivatives and the option to annuitize, {\it Insurance: Mathematics and Economics}, 29 (3): 299-318.

\smallskip \noindent \hangindent 20 pt Milevsky, M. A., S. D. Promislow, and V. R. Young (2005), Financial valuation of mortality risk via the instantaneous Sharpe ratio, working paper, Department of Mathematics, University of Michigan.

\smallskip \noindent \hangindent 20 pt Milevsky, M. A., S. D. Promislow, and V. R. Young (2007), Killing the law of large numbers: mortality risk premiums and the Sharpe ratio, {\it Journal of Risk and Insurance}, 73 (4): 673-686.

\smallskip \noindent \hangindent 20 pt Olshansky, O. J., B. A. Carnes, and C. Cassel (1990), In search of Methuselah: estimating the upper limits to human longevity, {\it Science}, 250: 634-640.

\smallskip \noindent \hangindent 20 pt Schrager, D. F. (2006), Affine stochastic mortality, {\it Insurance: Mathematics and Economics}, 38 (1): 81-97.

\smallskip \noindent \hangindent 20 pt Schweizer, M. (2001), A guided tour through quadratic hedging approaches, in {\it Option Pricing, Interest Rates and Risk Management}, E. Jouini, J. Cvitani\'c, and M. Musiela (editors), Cambridge University Press: 538-574.

\smallskip \noindent \hangindent 20 pt Walter, W. (1970), {\it Differential and Integral Inequalities}, Springer-Verlag, New York.

\smallskip \noindent \hangindent 20 pt Zariphopoulou, T., (2001), Stochastic control methods in asset pricing, {\it Handbook of \break Stochastic Analysis and Applications}, D. Kannan and V. Lakshmikantham (editors), Marcel Dekker, New York.

\bye